%% file: 00_main.tex
\documentclass[sigplan,10pt]{acmart}
\renewcommand\footnotetextcopyrightpermission[1]{}
\usepackage{microtype}
\usepackage{graphicx}
\usepackage{subfigure}
\usepackage{booktabs}
\usepackage{tikz}
\usetikzlibrary{arrows.meta,positioning,fit,calc}
\usepackage{amsmath}
\usepackage{xspace}
\usepackage{cleveref}
\usepackage{balance}
\usepackage{url}
\usepackage{enumitem}
\usepackage{listings}
\usepackage{xcolor}
\usepackage{amssymb}
\usepackage{pifont}
\usepackage{siunitx}
\usepackage{comment}
\usepackage{float}
\usepackage{hyperref}

\definecolor{CBlue}{HTML}{5B9BD5}
\definecolor{COrange}{HTML}{ED7D31}
\definecolor{CGreen}{HTML}{70AD47}
\definecolor{CRed}{HTML}{FF6B6B}
\definecolor{CGray}{HTML}{A9A9A9}
\definecolor{CPurple}{HTML}{7B68EE}
\definecolor{CTeal}{HTML}{2EC4B6}
\definecolor{CDkBlue}{HTML}{2F5597}

\newcommand{\sys}{Concordia\xspace}
\newcommand{\tech}{\sys}

\newcommand{\xiangyu}[1]{}

  {\begin{enumerate}[itemsep=-0pt, parsep=0pt, topsep=0pt, leftmargin=1pc]}
  {\end{enumerate}}

\crefformat{section}{\S#2#1#3}
\crefformat{subsection}{\S#2#1#3}
\crefformat{subsubsection}{\S#2#1#3}
\Crefformat{section}{Section~#2#1#3}
\Crefformat{subsection}{Section~#2#1#3}
\Crefformat{figure}{Figure~#2#1#3}
\crefformat{figure}{Figure~#2#1#3}

\begin{document}

\title{\sys: JIT-Compiled Persistent-Kernel Checkpointing for Fault-Tolerant LLM Inference}
\author{Yuhang Gan}
\affiliation{
  \institution{University of California Santa Cruz}
  \country{USA}
}
\email{ygan11@ucsc.edu}

\author{Yiwei Yang}
\affiliation{
  \institution{UC Santa Cruz}
  \country{USA}
}
\email{yyang363@ucsc.edu}

\author{Yuyi Li}
\affiliation{
  \institution{University of California, Davis}
  \country{USA}
}
\email{nylli@ucdavis.edu}

\author{Xiangyu Gao}
\affiliation{
  \institution{University of Washington}
  \country{USA}
}
\email{xiangyug@cs.washington.edu}

\author{Yichen Wang}
\affiliation{
  \institution{ByteDance}
  \country{USA}
}
\email{yichen.wang@bytedance.com}

\author{Rain Jiang}
\affiliation{
  \institution{ByteDance}
  \country{USA}
}
\email{rain.jiang@bytedance.com}

\author{Xiaoning Ding}
\affiliation{
  \institution{ByteDance}
  \country{USA}
}
\email{xiaoning.ding@bytedance.com}

\author{Andi Quinn}
\affiliation{
  \institution{University of California Santa Cruz}
  \country{USA}
}
\email{aquinn1@ucsc.edu}

\author{Chen Qian}
\affiliation{
  \institution{University of California Santa Cruz}
  \country{USA}
}
\email{cqian12@ucsc.edu}

\begin{abstract}
Long-running LLM agents keep valuable state resident on GPUs: KV caches, request schedulers, communication state, and sometimes online adapters.
Losing this state after a GPU or communicator failure can discard minutes to hours of work, yet existing recovery mechanisms either restart the whole serving stack or require application-specific checkpoint logic inside every attention and runtime component.
This paper argues that fault tolerance for such workloads needs a GPU-resident execution context: checkpoint hooks must run at device synchronization points, observe binary kernels that frameworks and libraries actually execute, and recover without putting the host CPU on the critical path.

We present \sys, a runtime that uses a device-resident persistent kernel as the substrate for fault-tolerant LLM inference.
\sys interposes on GPU module loading and supports PTX- and SASS-level instrumentation, allowing checkpoint and pause hooks to be inserted below framework code and library boundaries.
For each registered LLM state region, \sys JIT-compiles a specialized delta-checkpoint handler---for example, a KV-block scanner, adapter-page scanner, or recovery applier---and hot-swaps it into the persistent kernel's operator table.
The persistent kernel consumes a lock-free ring buffer of compute, checkpoint, append-log, and recovery tasks, so the same always-on executor triggers dirty-page detection, stages deltas, and appends committed records to a CPU-visible log in CXL memory or host DRAM.

\sys assumes a standard fail-stop model in which application kernels, GPU ranks, or communicators may fail, while the small persistent checkpoint worker is the trusted control loop until the device is declared lost; if the device disappears, recovery uses the last committed append-only log record.
\sys exploits the structure of LLM inference without requiring the application to log every KV update: base weights remain static, while KV-cache and adapter pages change sparsely and can be detected on the GPU at HBM bandwidth.
On RTX PRO 6000 Blackwell, GPU-side delta checkpointing is up to 219$\times$ faster than CPU-side page scanning, per-boundary checkpoint triggers avoid extra kernel launches, and a two-GPU recovery prototype restores service in about 1.5\,s rather than restarting NCCL and reloading the model.
\end{abstract}

\settopmatter{printfolios=true}
\maketitle
\pagestyle{plain}
\section{Introduction}
\label{sec:intro}
\input{./01_intro.tex}

\section{Background and Motivation}
\label{sec:background}
\input{./02_background.tex}

\section{Design}
\label{sec:method}
\input{./03_method.tex}

\section{Implementation}
\label{sec:implementation}
\input{./04_implement.tex}

\section{Evaluation}
\label{sec:evaluation}
\input{./05_eval.tex}

\section{Related Work}
\label{sec:related}
\input{./06_related.tex}

\section{Discussion}
\label{sec:discussion}
\input{./07_discuss.tex}

\section{Conclusion}
\label{sec:conclusion}
\input{./08_conclusion.tex}

\bibliographystyle{plain}
\bibliography{ref}

\end{document}

%% file: 01_intro.tex
Large Language Model (LLM) serving is moving from short, stateless requests to long-running agents that maintain large GPU-resident state across many turns, tool calls, and sometimes online adaptation~\cite{brown2020language,chowdhery2022palm,kojima2025lora}.
For these workloads, fault tolerance is no longer a background availability feature.
A failed GPU can destroy an active KV cache, scheduler state, in-flight collective, and adapter update accumulated over a long session.
Restarting the process, reloading model weights, and replaying the conversation is too slow, and often semantically impossible once tool calls have affected the outside world.

The obvious alternative is application-specific checkpointing.
For LLM inference, the dominant changing state is well structured: each decoded token appends KV entries, while base weights are static and LoRA adapters mutate only a small parameter subset.
A hand-written serving engine could log those KV writes directly.
This is an important baseline, but it is not a complete systems substrate.
Modern serving stacks combine PagedAttention allocators~\cite{kwon2023vllm}, fused attention libraries, generated Triton/CUDA kernels, NCCL collectives, framework graph compilers, and vendor libraries.
The recovery contract crosses module and binary boundaries: it must know when device work is quiescent, which physical cache blocks have changed, whether a collective can be safely bypassed, and where execution can resume.
Re-implementing this contract separately inside every framework path is brittle.

\sys takes a different position: fault tolerance for long-running LLM inference should be implemented below the framework, at the GPU binary/runtime boundary.
This makes the central technical problem different from traditional LLVM or x86 instrumentation.
GPU kernels execute under SIMT semantics, synchronize at CTA and collective boundaries, use device memory allocators whose physical layout is managed by the serving runtime, and are launched through a host driver path that may itself be stalled during recovery.
Instrumentation must therefore work on PTX and SASS kernels that the framework actually loads, while the checkpoint executor must already be resident on the GPU before a failure occurs.

The core mechanism in \sys is a device-resident \emph{persistent kernel}.
It runs for the lifetime of an LLM session, polls a lock-free ring buffer in host-mapped memory, and executes compute, checkpoint, communication, and recovery tasks without requiring a fresh host-launched CUDA kernel.
This persistent kernel is not introduced primarily as a faster replacement for CUDA Graphs on static inference.
CUDA Graphs remain effective for stable decode regions and are complementary to \sys.
The reason \sys needs persistence is fault tolerance: checkpoint hooks need a live device-side executor at NCCL boundaries; dirty-page detection needs to scan GPU memory at HBM bandwidth; and recovery needs a control path that does not depend on rebuilding the failed communicator before any GPU code can run.

The persistent executor enables three mechanisms that are useful only when co-designed.
First, \sys interposes on GPU module loading and instruments PTX/SASS code to insert cooperative pause and checkpoint hooks at kernel and collective boundaries.
Second, \sys JIT-compiles checkpoint handlers for the registered memory layout of each workload: a PagedAttention KV arena receives a block-table-aware scanner, LoRA adapters receive dense page scanners, and opaque mutable buffers receive shadow-compare scanners.
These handlers are hot-swapped into the persistent kernel's operator table and invoked as ring-buffer tasks.
This is transparent to the application while still exploiting LLM structure: base weights are registered as immutable regions, KV/cache allocators register their physical blocks, and adapter/optimizer regions are tracked separately.
Third, \sys persists the resulting deltas as an append-only recovery log in CPU-visible CXL memory or host DRAM.
The design is analogous to Redis AOF: every committed state mutation is appended to a sequential log, and periodic compaction rewrites the log into a smaller base snapshot plus recent deltas.
Dispatch acceleration and JIT amortization are consequences of the same persistent context, not independent claims.

\sys assumes the persistent checkpoint worker is a trusted control loop.
It may observe failures in application kernels, GPU ranks, or NCCL communicators; it is not designed to recover from an independent software crash of the persistent worker itself.
If the whole device is lost, recovery starts from the last committed AOF record in CXL/DRAM on a replacement GPU.

This framing also clarifies what \sys does not claim.
For a static single-model decode benchmark, a carefully engineered CUDA Graph path may outperform eager PyTorch and can be used alongside \sys.
For a serving engine willing to modify every attention kernel and allocator, explicit KV logging can reduce checkpoint data movement further than transparent page tracking.
\sys targets the harder deployment point: unmodified or partially modified GPU binaries, dynamic serving paths, and failure recovery across framework/library boundaries.

In summary, this paper makes the following contributions:
\begin{itemize}[nosep]
\item We identify persistent-kernel execution as the missing substrate for transparent LLM fault tolerance: without a live device-side executor, checkpointing and recovery fall back to host launches, CPU page scans, and full communicator restarts.

\item We design \sys, a GPU runtime that combines PTX/SASS instrumentation, a lock-free persistent executor, JIT-compiled GPU-side delta checkpoint handlers, and append-only CXL/DRAM recovery logging under one recovery contract.

\item We show how \sys exploits LLM state structure without requiring application-specific KV logging: immutable weight regions, mutable KV/cache pages, and adapter pages are registered and diffed on the GPU at HBM bandwidth.

\item We evaluate \sys on Blackwell GPUs and heterogeneous targets, showing up to 219$\times$ faster delta checkpointing than CPU-side page scanning, sub-microsecond checkpoint trigger submission, low persistent-kernel SM footprint, and second-scale recovery in a two-GPU prototype.
\end{itemize}

%% file: 02_background.tex
\sys is motivated by a fault-tolerance problem rather than by launch overhead alone.
Long-running LLM inference keeps three classes of state on GPUs:
(1) immutable model weights,
(2) append-heavy KV-cache and scheduler state, and
(3) small mutable regions such as LoRA adapters and optimizer state.
The first class is large but rarely changes; the second and third classes change frequently and determine whether a session can resume after a failure.
The system question is where this state should be observed and checkpointed.

\subsection{Why Application-Level KV Logging Is Not Enough}

For a single attention implementation, the most efficient checkpoint is clear: record the KV slice appended by each decoded token and replay it during recovery.
\sys does not dispute this.
The difficulty is that production serving stacks do not expose one uniform KV write path.
PagedAttention~\cite{kwon2023vllm} maps logical tokens onto dynamically allocated physical cache blocks; fused attention kernels may update multiple cache regions; LoRA or test-time adaptation changes separate parameter pages; and communication libraries mutate collective buffers outside the model code.
An application-level logger must be threaded through all of these paths and kept consistent with every generated kernel and vendor library version.

\sys instead uses page-level tracking as a transparent recovery contract.
The serving runtime can still provide semantic hints, such as registering PagedAttention physical KV blocks and marking base weights immutable, but correctness does not depend on hand-logging every token update.
This is the reason \sys instruments GPU binaries rather than only compiler IR from the application source: the relevant writes may occur in PTX/SASS kernels emitted by Triton, NVRTC, vendor libraries, or framework code.

\subsection{Why the Checkpoint Executor Must Be Persistent}

Host-mediated checkpointing has two costs.
First, the host must launch the checkpoint kernel or copy operation at every checkpoint boundary.
This overhead is small for occasional snapshots, but it matters when checkpoints are tied to decode or NCCL boundaries.
Second, if dirty-page detection runs on the CPU, the host must either copy and scan large GPU regions or rely on application-specific logs.
The former wastes bandwidth and CPU time; the latter gives up transparency.

A persistent GPU executor changes the control path.
Once launched, it can receive a checkpoint task through a ring buffer, scan registered GPU regions at HBM bandwidth, stage dirty pages, and update metadata without a new CUDA launch.
It also remains available while the host is handling agent orchestration or while a communicator is being repaired.
This is different from CUDA Graphs~\cite{NVBLOGGraphs2019,CTGraphLaunch2024}.
Graphs reduce launch overhead for replayable DAGs and are useful for stable decode regions, but they do not by themselves provide a live recovery executor, binary-level pause hooks, or dirty-page scanning across framework/library boundaries.

\subsection{Why Append Deltas to CXL or DRAM}

Once dirty state is discovered on the GPU, the recovery target should not be another GPU-only shadow copy.
HBM is scarce, and a failed GPU may take its local memory with it.
\sys instead treats deltas as a sequential recovery stream, similar to an append-only file (AOF) in Redis.
Each checkpoint boundary appends a compact record containing a region ID, version, dirty page list, payload, and commit marker to CPU-visible storage.
The natural targets are host DRAM and CXL memory pools~\cite{cxl2024specification}, which provide byte-addressable capacity close enough to the GPU node to serve as a low-latency recovery log.

This log has two advantages over periodic full snapshots.
First, it matches the mutation pattern of LLM inference: KV-cache and adapter updates are append-heavy and sparse.
Second, it separates durability from HBM capacity.
Recent records can stay in DRAM/CXL for fast restart, while a background compactor periodically rewrites the log into a consolidated base snapshot plus a short suffix of deltas.
Recovery replays the latest base snapshot and AOF suffix onto a replacement GPU.

\subsection{Motivating Experiment: Host-Side Dirty Detection}
\label{sec:motivating}

We quantify the cost of host-side dirty detection because it is the operation \sys moves most directly into the persistent GPU executor.
The experiment simulates a per-token KV-cache update: a single contiguous 4\,KB page is modified in a GPU memory region of 16--256\,MB.
This is intentionally a page-level transparent checkpoint, not an application-specific KV logger.

\begin{figure}[t]
    \centering
    \includegraphics[width=\columnwidth]{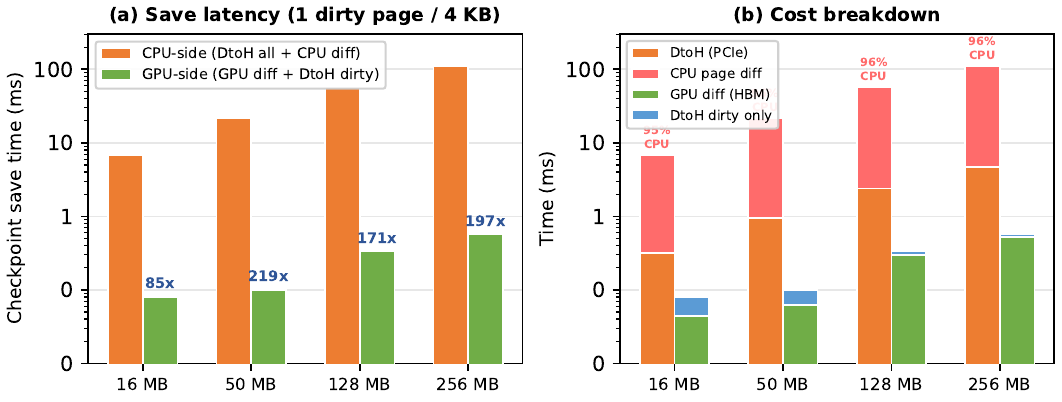}
    \caption{Motivating experiment on RTX PRO 6000 Blackwell.
    (a)~Checkpoint save latency for a single dirty page (4\,KB), simulating a sparse KV-cache update.
    CPU-side delta checkpointing transfers and scans the full region; GPU-side delta checkpointing scans at HBM bandwidth and transfers only dirty pages.
    (b)~Cost breakdown: host page comparison dominates CPU-side transparent checkpointing, while GPU-side diffing is bounded by HBM bandwidth.}
    \label{fig:motivating}
\end{figure}

\textbf{CPU-full} copies the entire GPU region to pinned host memory with \texttt{cuMemcpyDtoH()}.
\textbf{CPU-delta} copies the region and compares it against a host shadow copy at 4\,KB granularity, storing only dirty pages.
\textbf{GPU-delta} compares current data with a GPU-resident shadow copy on device, then transfers only dirty pages to host memory.

The results in \Cref{fig:motivating} show three points.

\textit{(1)~Transparent CPU-side diffing is dominated by host scanning.}
For a 256\,MB region, the device-to-host copy takes 4.72\,ms, while page comparison takes 106.65\,ms in our prototype.
This comparison path uses Python/NumPy, so the exact multiplier is not a claim about an optimized checkpoint library.
The underlying scaling remains the problem: a transparent CPU diff must read the whole region from host memory even when only one page changed.

\textit{(2)~GPU-side diffing matches the hardware locality of the data.}
The GPU comparison kernel reads current and shadow buffers at HBM bandwidth, taking 0.04--0.53\,ms for 16--256\,MB, and transfers only the dirty 4\,KB page.
The measured end-to-end latency is 0.08--0.56\,ms, up to 219$\times$ faster than our CPU-side transparent prototype.
The important point is not the exact multiplier, but the cost model:
CPU-side transparent diffing scales with total region size at CPU memory bandwidth, whereas GPU-side diffing scales with total region size at HBM bandwidth plus dirty bytes over PCIe.

\textit{(3)~Checkpoint triggers should not require new host launches.}
If each checkpoint is a separate \texttt{cudaLaunchKernel}, the host remains on the recovery critical path.
The persistent executor turns checkpointing into a task descriptor submitted to an already-running kernel.
This same mechanism handles dispatch and communication tasks, but in \sys those are supporting roles for the recovery substrate.

These observations lead to \sys's design: instrument GPU binaries to expose safe checkpoint/pause points, keep a persistent device-side executor alive for the whole session, JIT-compile checkpoint handlers specialized to the registered memory layout, and append committed deltas to a CXL/DRAM recovery log.

%% file: 03_method.tex
\pgfdeclarelayer{background}
\pgfsetlayers{background,main}

\sys is organized around one recovery invariant: before a failure, the GPU must already contain a small executor capable of observing safe points, running checkpoint code, and appending recovery records without asking the host to launch new kernels.
The device-resident \emph{persistent kernel} provides this executor (\cref{sec:persistent}).
PTX/SASS instrumentation inserts cooperative pause and checkpoint hooks into the kernels that frameworks and libraries actually load, while runtime JIT compilation specializes delta-checkpoint handlers for the registered LLM memory layout (\cref{sec:portable}).
GPU-side delta checkpointing then runs as persistent-kernel tasks at kernel and NCCL collective boundaries, scanning mutable LLM regions at HBM bandwidth and appending dirty-page records to a CXL/DRAM recovery log (\cref{sec:fault}).
Dispatch acceleration, operator hot-swap, and optional portable recovery follow from this substrate, but the design is driven by checkpoint/replay.

\subsection{Persistent Kernel Runtime}
\label{sec:persistent}
The foundation of \sys is a persistent kernel that continuously processes work from a shared queue.
In the fault-tolerance configuration evaluated in this paper, the executor reserves one resident worker block on the GPU, giving the runtime a live device-side control path at 0.53\% SM footprint on RTX PRO 6000.
For pure micro-dispatch experiments the same code can be launched with more worker blocks, but the recovery path does not require one block per SM.
This distinction matters: \sys relies on persistence, not on occupying the whole device.

\paragraph{Fault model.}
\sys targets fail-stop faults in application kernels, GPU ranks, and collective communication.
The persistent checkpoint worker is treated as a trusted control loop: it is small, launched before application work, monitored by a heartbeat, and assumed not to fail independently of the device.
If the worker heartbeat stops, \sys treats the GPU as lost and recovers from the last committed append-only log record in CXL/DRAM.
The design does not handle Byzantine corruption of the persistent worker or arbitrary silent data corruption inside HBM.

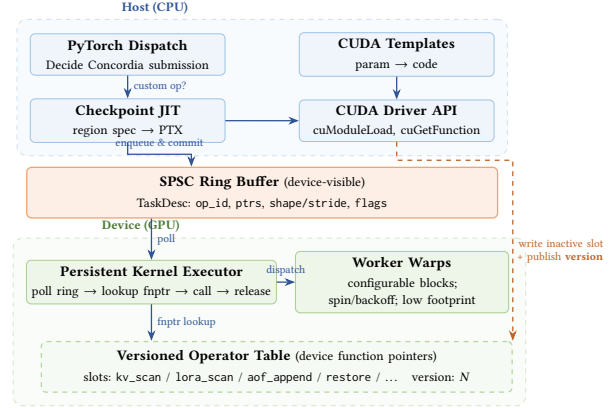
\begin{figure}[t]
    \centering
    \resizebox{\columnwidth}{!}{%
    \begin{tikzpicture}[
        >=Stealth,
        node distance=0.6cm and 0.8cm,
        hostbox/.style={draw=CBlue!60, rounded corners=3pt, fill=CBlue!10,
                        minimum height=0.9cm, text width=4.0cm,
                        align=center, font=\small},
        ringbox/.style={draw=COrange!70, thick, rounded corners=3pt, fill=COrange!12,
                        minimum height=1.1cm, text width=10.0cm,
                        align=center, font=\small},
        execbox/.style={draw=CGreen!60, rounded corners=3pt, fill=CGreen!12,
                        minimum height=0.9cm, align=center, font=\small},
        opbox/.style={draw=CGreen!50, dashed, thick, rounded corners=3pt, fill=CGreen!6,
                      minimum height=1.1cm, text width=10.0cm,
                      align=center, font=\small},
        zone/.style={draw, gray, dashed, rounded corners=6pt, inner sep=8pt},
        lbl/.style={font=\footnotesize\bfseries, text=gray!70!black},
        arr/.style={->, thick, CDkBlue},
        darr/.style={->, thick, dashed, COrange!80!black},
    ]

    \node[hostbox] (dispatch)
        {\textbf{PyTorch Dispatch}\\[1pt]
         {\footnotesize Decide Concordia submission}};
    \node[hostbox, right=1.6cm of dispatch] (templates)
        {\textbf{CUDA Templates}\\[1pt]
         {\footnotesize param $\rightarrow$ code}};

    \node[hostbox, below=0.5cm of dispatch] (nvrtc)
        {\textbf{Checkpoint JIT}\\[1pt]
         {\footnotesize region spec $\rightarrow$ PTX}};
    \node[hostbox, below=0.5cm of templates] (driver)
        {\textbf{CUDA Driver API}\\[1pt]
         {\footnotesize cuModuleLoad, cuGetFunction}};

    \begin{pgfonlayer}{background}
        \node[zone, fill=CBlue!4, draw=CBlue!30,
              fit=(dispatch)(templates)(nvrtc)(driver),
              label={[lbl, text=CDkBlue]above left:Host (CPU)}] (hostzone) {};
    \end{pgfonlayer}

    \draw[arr] (dispatch) -- node[right, font=\scriptsize]{custom op?} (nvrtc);
    \draw[arr] (templates) -- (driver);
    \draw[arr] (nvrtc) -- (driver);

    \node[ringbox, below=1.0cm of $(nvrtc)!0.5!(driver)$] (ring)
        {\textbf{SPSC Ring Buffer} {\footnotesize (device-visible)}\\[2pt]
         {\footnotesize TaskDesc:\ \texttt{op\_id},
          \texttt{ptrs}, \texttt{shape/stride}, \texttt{flags}}};

    \node[execbox, text width=5.2cm,
          below=0.9cm of ring.south west, anchor=north west] (executor)
        {\textbf{Persistent Kernel Executor}\\[1pt]
         {\footnotesize poll ring $\rightarrow$ lookup fnptr
          $\rightarrow$ call $\rightarrow$ release}};
    \node[execbox, text width=4.4cm,
          right=0.4cm of executor] (warps)
        {\textbf{Worker Warps}\\[1pt]
         {\footnotesize configurable blocks; spin/backoff; low footprint}};

    \node[opbox, below=0.7cm of $(executor.south)!0.5!(warps.south)$] (optable)
        {\textbf{Versioned Operator Table}
         {\footnotesize (device function pointers)}\\[2pt]
         {\footnotesize slots:\ \texttt{kv\_scan}\ /\ \texttt{lora\_scan}\
          /\ \texttt{aof\_append}\ /\ \texttt{restore}\ /\ \dots
          \quad version:\ \textit{N}}};

    \begin{pgfonlayer}{background}
        \node[zone, fill=CGreen!4, draw=CGreen!30,
              fit=(executor)(warps)(optable),
              label={[lbl, text=CGreen!70!black]above left:Device (GPU)}] (devzone) {};
    \end{pgfonlayer}

    \coordinate (rinktgt) at ($(ring.north west)!0.35!(ring.north east)$);
    \draw[arr] (nvrtc.south) -- ++(0,-0.25)
        -| node[pos=0.25, above, font=\scriptsize]
           {enqueue \& commit} (rinktgt);

    \draw[arr] (ring.south -| executor.north) -- (executor.north)
        node[midway, right, font=\scriptsize]{poll};

    \draw[arr] (executor.south) -- (executor.south |- optable.north)
        node[midway, right, font=\scriptsize]{fnptr lookup};

    \draw[arr] (executor.east) -- (warps.west)
        node[midway, above, font=\scriptsize]{dispatch};

    \draw[darr] (driver.south) -- ++(0,-0.4)
        -| node[pos=0.75, right, font=\scriptsize, text width=2.8cm, align=left]
           {write inactive slot\\+ publish \textbf{version}}
        (optable.north east);

    \end{tikzpicture}%
    }
\caption{\sys persistent kernel architecture: a ring buffer connects the host submission path to persistent device threads, with a dynamically updatable operator table for JIT-compiled checkpoint and recovery handlers. Solid arrows show the steady-state data path; the dashed arrow shows the hot-swap injection path for new handlers.}
    \label{fig:persistent-arch}
\end{figure}

The persistent runtime consists of three components.
A lock-free ring buffer in device-mapped memory connects host submissions to device execution: the host enqueues compact task descriptors (64--128~bytes including operator ID, tensor pointers, dimensions, and control flags) with store-release semantics, while device threads poll a read cursor with load-acquire semantics to ensure visibility.
A single persistent kernel launches at process startup and remains resident; each worker warp claims work atomically from the ring, dispatches through a function pointer table, and returns for the next task as shown in Fig.~\ref{fig:persistent-arch}.
The operator table is a device-resident array indexed by operator ID.
Checkpoint handlers are compiled via NVRTC~\cite{NVRTC} from region specifications, loaded through the CUDA Driver API~\cite{CudaDriverModule}, written into an inactive slot, and made visible by flipping a version counter, enabling hot-swapping without service interruption.

The host-side path is streamlined for minimal overhead:

\begin{lstlisting}
// Enqueue atomically
TaskDescriptor desc = {
  .op_id = OP_ADD,
  .input_a = a.data(), .input_b = b.data(),
  .output = c.data(), .size = a.size()
};
uint32_t slot = ring_buffer.acquire_slot();
ring_buffer.write(slot, desc);
ring_buffer.commit(slot); // store-release
\end{lstlisting}

Device-side dispatch treats checkpoint and recovery as first-class tasks rather than special host-launched kernels:

\begin{lstlisting}
while (true) {
    TaskDescriptor desc;
    if (!ring_buffer.poll_acquire(&desc)) {
        backoff_or_yield();
        continue;
    }

    Handler fn = operator_table.load(desc.op_id,
                                     desc.version);
    switch (desc.kind) {
    case TASK_COMPUTE:
        fn(desc);
        break;
    case TASK_DELTA_CKPT:
        Delta d = fn.scan_dirty(desc.region);
        aof_append(desc.epoch, d);      // CXL/DRAM log
        break;
    case TASK_RESTORE:
        fn.apply(aof_read(desc.epoch));
        break;
    }
    ring_buffer.complete_release(desc.seq);
}
\end{lstlisting}

This achieves host-side task submission latencies under 100~ns.
End-to-end completion latency also includes GPU polling, operator execution, and system-scope completion fences; \cref{sec:evaluation} reports both.

\sys integrates with PyTorch through dispatch interposition at the autograd engine level.
A dispatcher hook evaluates whether each operation is a candidate for ring buffer submission based on operation type (favoring element-wise ops, small reductions, cache updates, and checkpoint triggers), tensor size, and current load.
Eligible operations route through \sys; ineligible ones proceed through PyTorch's normal path.
This hybrid approach is intended to coexist with CUDA Graph capture for stable decode regions rather than replace it.

\Cref{tab:gpuos_api} summarizes the host-side API exposed by \sys and is used by the PyTorch interposition layer to configure capacity, fusion, yielding, and liveness checks during deployment.

\begin{table}[t]
\centering
\caption{\sys host-side runtime API.}
\label{tab:gpuos_api}
\begin{tabular}{ll}
\toprule
\textbf{Function} & \textbf{Description} \\
\midrule
init(capacity, tpb) & Initialize runtime, allocate queue \\
fuse() & Fuse consecutive operations \\
set\_yield\_every(n) & Control yield policy (0 = never) \\
peek\_queue() & Query queue state \\
worker\_alive() & Check if persistent kernel runs \\
shutdown() & Signal exit, free resources \\
\bottomrule
\end{tabular}
\end{table}

The persistent kernel's continuous device presence extends beyond operator dispatch.
Because it maintains direct access to device memory and a steady execution context, \sys uses the ring buffer to enqueue checkpoint scans, AOF appends, restore operations, and optional GPU-initiated network transfers without returning to the host.
This makes the host CPU stateless with respect to the inference data path: the GPU autonomously dispatches checkpoint handlers, appends committed deltas, and can initiate inter-node communication, while the CPU handles only initialization and agent-level orchestration such as tool calls.

\subsection{PTX/SASS Instrumentation and JIT Checkpoint Handlers}
\label{sec:portable}

Because the persistent kernel never exits, it provides a stable device-resident context into which \sys can install checkpoint code after model loading.
\sys uses PTX/SASS instrumentation to discover safe points and memory regions, then JIT-compiles checkpoint handlers specialized to the registered layout.
The JIT target is the persistent kernel's operator table, not a separate application kernel launch.
For example, a PagedAttention region produces a handler that walks the physical block table and dirty bitmap; a LoRA region produces a dense page scanner; an opaque region produces a shadow-compare scanner; and each region gets a matching restore/applier handler.
When recovery must move across GPU architectures, the same instrumented state can optionally be lowered through CTX, but cross-architecture portability is a recovery extension rather than the main purpose of the JIT.

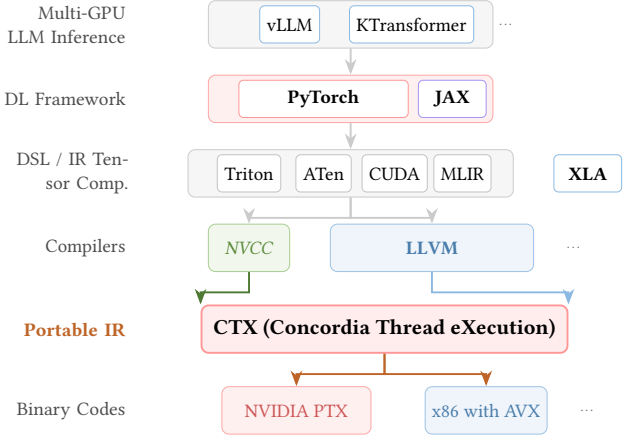
\begin{figure}[t]
    \centering
    \resizebox{\columnwidth}{!}{%
    \begin{tikzpicture}[
        >=Stealth,
        layerbox/.style={draw=#1!60, fill=#1!10, rounded corners=3pt,
                         minimum height=0.7cm, align=center, font=\small,
                         inner sep=4pt},
        itembox/.style={draw=#1!70, fill=white, rounded corners=2pt,
                        minimum height=0.55cm, align=center, font=\footnotesize,
                        inner sep=3pt},
        irbox/.style={draw=CRed!70, fill=CRed!12, rounded corners=3pt,
                      minimum height=0.7cm, align=center, font=\small\bfseries,
                      inner sep=5pt, thick},
        arr/.style={->, thick, CDkBlue},
        lbl/.style={font=\footnotesize, text=gray!60!black, anchor=east},
    ]
    \node[layerbox=CGray, minimum width=4.2cm] (vllm) at (0, 0)
        {};
    \node[itembox=CBlue] at (-0.9, 0) {vLLM};
    \node[itembox=CBlue] at (0.9, 0) {KTransformer};
    \node[font=\tiny, text=CGray!70!black] at (2.3, 0) {\dots};
    \node[lbl, text width=2cm, align=right] at (-3.2, 0) {Multi-GPU LLM Inference};

    \node[layerbox=CRed, minimum width=4.2cm] (pytorch) at (0, -1.1) {};
    \node[itembox=CRed, minimum width=2.5cm] at (-0.4, -1.1) {\textbf{PyTorch}};
    \node[itembox=CPurple, minimum width=1.0cm] at (1.5, -1.1) {\textbf{JAX}};
    \node[lbl, text width=2cm, align=right] at (-3.2, -1.1) {DL Framework};

    \node[layerbox=CGray, minimum width=4.8cm] (dsl) at (0, -2.2) {};
    \node[itembox=CGray, font=\footnotesize] at (-1.5, -2.2) {Triton};
    \node[itembox=CGray, font=\footnotesize] at (-0.4, -2.2) {ATen};
    \node[itembox=CGray, font=\footnotesize] at (0.65, -2.2) {CUDA};
    \node[itembox=CGray, font=\footnotesize] at (1.65, -2.2) {MLIR};
    \node[itembox=CBlue, minimum width=1.0cm] at (3.5, -2.2) {\textbf{XLA}};
    \node[lbl, text width=2cm, align=right] at (-3.2, -2.2) {DSL / IR Tensor Comp.};

    \node[layerbox=CGreen, minimum width=1.2cm] (nvcc) at (-1.5, -3.3) {};
    \node[font=\footnotesize, text=CGreen!80!black] at (-1.5, -3.3) {\textit{NVCC}};
    \node[layerbox=CBlue, minimum width=3.0cm] (llvm) at (1.2, -3.3) {};
    \node[font=\footnotesize\bfseries, text=CBlue!80!black] at (1.2, -3.3) {LLVM};
    \node[font=\tiny, text=CGray!70!black] at (3.3, -3.3) {\dots};
    \node[lbl] at (-3.2, -3.3) {Compilers};

    \node[irbox, minimum width=3.5cm] (ctx) at (0.5, -4.5)
        {CTX (Concordia Thread eXecution)};
    \node[lbl, text=COrange!80!black, font=\footnotesize\bfseries, text width=2cm, align=right] at (-3.2, -4.5) {Portable IR};

    \node[layerbox=CRed, minimum width=2.2cm] (ptx) at (-0.8, -5.7) {};
    \node[font=\footnotesize, text=CRed!80!black] at (-0.8, -5.7) {NVIDIA PTX};
    \node[layerbox=CBlue, minimum width=1.8cm] (x86) at (2.0, -5.7) {};
    \node[font=\footnotesize, text=CBlue!80!black] at (2.0, -5.7) {x86 with AVX};
    \node[font=\tiny, text=CGray!70!black] at (3.5, -5.7) {\dots};
    \node[lbl, text width=2cm, align=right] at (-3.2, -5.7) {Binary Codes};

    \draw[arr, CGray!60] (vllm.south) -- (pytorch.north);
    \draw[arr, CGray!60] (pytorch.south) -- (dsl.north);
    \draw[arr, CGray!60] (dsl.south) -- ++(0, -0.3) -| (nvcc.north);
    \draw[arr, CGray!60] (dsl.south) -- ++(0, -0.3) -| (llvm.north);
    \draw[arr, CGreen!70!black] (nvcc.south) -- ++(0, -0.3) -| (ctx.north west);
    \draw[arr, CBlue!70] (llvm.south) -- ++(0, -0.3) -| (ctx.north east);
    \draw[arr, COrange!80!black, thick] (ctx.south) -- ++(0, -0.3) -| (ptx.north);
    \draw[arr, COrange!80!black, thick] (ctx.south) -- ++(0, -0.3) -| (x86.north);

    \end{tikzpicture}%
    }
    \caption{\sys instrumentation and recovery pipeline. CUDA/PTX/SASS modules are instrumented with pause/checkpoint hooks; registered memory layouts drive JIT generation of persistent-kernel checkpoint handlers. CTX lowering is used only when recovery crosses architectures.}
    \label{fig:portable-arch}
\end{figure}

As shown in Fig.~\ref{fig:portable-arch}, \sys first instruments the GPU code that the framework actually loads.
For PTX modules, \sys rewrites the PTX before driver JIT compilation.
For precompiled cubins or vendor libraries where PTX is unavailable, \sys uses SASS-level patching to redirect selected entry, exit, and barrier-adjacent points through trampoline stubs.
The PTX path is preferred because symbolic registers and memory spaces are still explicit; the SASS path is necessary for binary-only kernels and requires architecture-specific decoding, relocation, and register-liveness constraints.
This is why the problem differs from LLVM or x86 instrumentation: GPU instrumentation must preserve SIMT reconvergence, CTA barriers, memory-space qualifiers, and system-scope ordering while running inside a driver-controlled launch model.

\paragraph{Checkpoint-handler JIT.}
For every registered region, \sys builds a compact region specification: base address, page size, page count, mutability class, optional allocator metadata pointers, AOF record format, and restore policy.
The host JIT emits a CUDA template specialized to that specification and installs the resulting device function into the persistent kernel's operator table.
Specialization removes branches from the hot checkpoint path: KV handlers know whether they should read a dirty-block bitmap, adapter handlers know the dense page range, and opaque handlers know the shadow-buffer address.
The same JIT also emits the corresponding AOF append and restore handlers, so checkpoint and recovery use symmetric code paths.

\paragraph{Optional CTX lowering.}
When recovery stays on the same GPU architecture, the JIT-compiled checkpoint handlers and the AOF log are sufficient.
When recovery crosses architectures, \sys normalizes instrumented code into CTX (Concordia Thread eXecution), a compact assembly that records thread/block indices, memory spaces, barriers, predication, and explicit safe points.
We extend LLVM~22.0 with a custom backend (CTXTarget) that emits CTX assembly from instrumented IR.
The frontend remaps CUDA builtins to abstract intrinsics (\texttt{llvm.concordia.barrier}, \texttt{llvm.concordia.ld.shared}), and SASS hooks attach metadata that reconstructs the corresponding CTX-level state at safe points.
At runtime, CTX modules translate to the recovery target.

\paragraph{Per-target lowering.}
For NVIDIA/PTX, CTX instructions map to PTX via NVVM: \texttt{GET\_GLOBAL\_ID} becomes a sequence reading \texttt{ctaid} and \texttt{tid} registers; predicated blocks map to \texttt{@predicate}-guarded instructions; the result is JIT-compiled to cubin via \texttt{ptxas}.
For AMD/ROCm, CTX lowers through AMD Comgr using appropriate storage classes (\texttt{CrossWorkgroup} for global, \texttt{Workgroup} for shared) and \texttt{OpControlBarrier} for workgroup-scope barriers, with predication realized through structured \texttt{OpSelectionMerge}/\texttt{OpBranchConditional}.
For Intel/SPIR-V, the output feeds Level Zero compilation.
For Tenstorrent/TOSA, CTX maps to TOSA MLIR tensors: scalar registers become 0-D tensors, vector registers become 1-D tensors; arithmetic uses \texttt{tosa.add}/\texttt{tosa.mul}; synchronization markers are handled by the TT-MLIR backend during lowering.

\paragraph{SIMT--MIMD reconciliation.}
The deepest challenge is reconciling SIMT and MIMD execution models. SIMT GPUs execute threads in lockstep warps with implicit synchronization, whereas Tenstorrent's architecture exposes explicit DMA and scratchpads without a warp scheduler. \sys employs two strategies. In vectorized warp emulation, a warp's threads are mapped onto a single core's vector unit to execute SIMD-width operations per cycle. In multi-core partitioning, a block's warps are distributed across multiple Tensix cores with explicit inter-core synchronization at barrier points. The runtime selects between these modes based on kernel characteristics: regular, vectorizable kernels use the former, while irregular kernels with divergence benefit from the latter.

\paragraph{Cooperative pause and resume.}
\sys does not attempt arbitrary instruction-level preemption.
At instrumented safe points, code checks a \texttt{pause\_flag} in device memory.
When set, threads write live register values, program-counter labels, CTA coordinates, and memory-region version numbers into a CTX state buffer.
On the target device, a resume kernel reconstructs the CTX state, remaps base addresses, restores registers through target-specific mechanisms, and jumps to the saved basic block.
The persistent executor coordinates this protocol and keeps the checkpoint metadata live while the host allocates a replacement device.
For kernels without safe points, \sys falls back to boundary checkpointing at kernel completion.

\subsection{GPU-Side Delta Checkpointing at NCCL Boundaries}
\label{sec:fault}

The persistent kernel's ring buffer provides a natural mechanism for GPU-side delta checkpointing at NCCL collective boundaries.
Rather than launching separate checkpoint kernels from the host, \sys executes dirty-page detection as a task within the already-resident persistent kernel---the same executor that handles operator dispatch and network initiation---achieving zero-overhead checkpoint triggers.
Combined with enhanced NCCL error handling and dynamic resource management, this enables rapid fault recovery for distributed LLM inference, directly motivated by the 85--219$\times$ speedup of GPU-side over CPU-side checkpointing demonstrated in \cref{sec:motivating}.

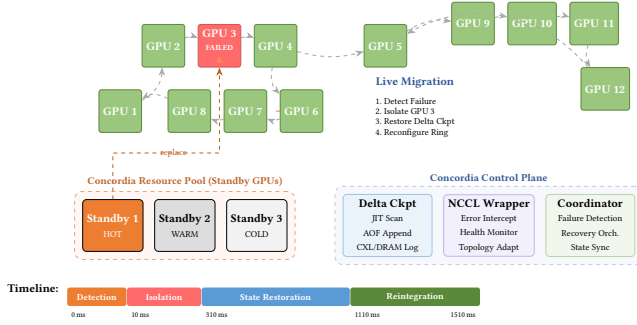
\begin{figure}[t]
\centering
\resizebox{\columnwidth}{!}{%
\begin{tikzpicture}[
    >=Stealth,
    node distance=0.4cm and 0.5cm,
    gpu/.style={draw=CGreen!80!black, fill=CGreen!70, text=white, font=\small\bfseries,
                rounded corners=2pt, minimum width=1.15cm, minimum height=1.15cm,
                align=center},
    failgpu/.style={draw=CRed!80!black, fill=CRed, text=white, font=\small\bfseries,
                    rounded corners=2pt, minimum width=1.15cm, minimum height=1.15cm,
                    align=center},
    standby/.style={draw, rounded corners=3pt, minimum width=1.5cm,
                    minimum height=1.4cm, align=center, font=\small},
    cpbox/.style={draw, rounded corners=3pt, minimum width=2.4cm,
                  minimum height=1.6cm, align=center, font=\small},
    zone/.style={draw, dashed, rounded corners=6pt, inner sep=6pt},
    phaselbl/.style={font=\scriptsize\bfseries, text=white, minimum height=0.5cm,
                     rounded corners=2pt, inner sep=3pt},
    arr/.style={->, thick, dashed, CGray},
    lbl/.style={font=\footnotesize\bfseries},
]

\node[gpu] (g2) at (1.5, 0) {GPU 2};
\node[failgpu, right=0.35cm of g2] (g3) {GPU 3\\[-1pt]{\tiny FAILED}\\[-2pt]{\tiny\color{yellow!90!black}$\triangle$}};
\node[gpu, right=0.35cm of g3] (g4) {GPU 4};
\node[gpu, below left=0.6cm and 0.0cm of g2] (g1) {GPU 1};
\node[gpu, below=0.6cm of g2, xshift=0.7cm] (g8) {GPU 8};
\node[gpu, right=0.35cm of g8] (g7) {GPU 7};
\node[gpu, right=0.35cm of g7] (g6) {GPU 6};

\node[gpu, right=1.8cm of g4] (g5) {GPU 5};
\node[gpu, right=1.2cm of g5, yshift=0.6cm] (g9) {GPU 9};
\node[gpu, right=0.35cm of g9] (g10) {GPU 10};
\node[gpu, right=0.35cm of g10] (g11) {GPU 11};
\node[gpu, below=0.6cm of g11, xshift=0.3cm] (g12) {GPU 12};

\draw[arr, bend left=20] (g2) to (g3);
\draw[arr, bend left=20] (g3) to (g4);
\draw[arr, bend right=30] (g4) to (g6);
\draw[arr, bend left=20] (g6) to (g7);
\draw[arr, bend left=20] (g7) to (g8);
\draw[arr, bend right=30] (g8) to (g1);
\draw[arr, bend right=30] (g1) to (g2);
\draw[arr, bend right=15] (g4) to (g5);
\draw[arr, bend left=20] (g5) to (g9);
\draw[arr, bend left=15] (g9) to (g10);
\draw[arr, bend left=15] (g10) to (g11);
\draw[arr, bend right=25] (g11) to (g12);
\draw[arr, bend right=40] (g12) to (g5);

\node[font=\small\bfseries, text=CDkBlue, below=0.15cm of g5, xshift=0.8cm] (miglab) {Live Migration};
\node[font=\scriptsize, text=black, below=0.0cm of miglab, anchor=north, align=left] {
  1.~Detect Failure\\
  2.~Isolate GPU~3\\
  3.~Restore Delta Ckpt\\
  4.~Reconfigure Ring};

\draw[CRed, thick, dashed] (g6.west) -- ++(-0.15,0);

\node[standby, fill=COrange, text=white,
      below=1.8cm of g1, xshift=-0.2cm] (sb1)
    {\textbf{Standby 1}\\{\scriptsize HOT}};
\node[standby, fill=CGray!40, right=0.3cm of sb1] (sb2)
    {\textbf{Standby 2}\\{\scriptsize WARM}};
\node[standby, fill=CGray!15, right=0.3cm of sb2] (sb3)
    {\textbf{Standby 3}\\{\scriptsize COLD}};

\begin{pgfonlayer}{background}
    \node[zone, fill=COrange!6, draw=COrange!70,
          fit=(sb1)(sb2)(sb3),
          label={[lbl, text=COrange!80!black]above:Concordia Resource Pool (Standby GPUs)}] (respool) {};
\end{pgfonlayer}

\draw[->, thick, COrange!80!black, dashed] (sb1.north) -- ++(0, 1.25) -| (g3.south)
    node[pos=0.20, right, font=\scriptsize]{replace};

\node[cpbox, fill=CBlue!12, draw=CBlue!60,
      right=1.5cm of sb3, yshift=0.0cm] (ckpt)
    {\textbf{Delta Ckpt}\\{\scriptsize JIT Scan}\\{\scriptsize AOF Append}\\{\scriptsize CXL/DRAM Log}};
\node[cpbox, fill=CPurple!10, draw=CPurple!50,
      right=0.3cm of ckpt] (nccl)
    {\textbf{NCCL Wrapper}\\{\scriptsize Error Intercept}\\{\scriptsize Health Monitor}\\{\scriptsize Topology Adapt}};
\node[cpbox, fill=CGreen!10, draw=CGreen!50,
      right=0.3cm of nccl] (coord)
    {\textbf{Coordinator}\\{\scriptsize Failure Detection}\\{\scriptsize Recovery Orch.}\\{\scriptsize State Sync}};

\begin{pgfonlayer}{background}
    \node[zone, fill=CDkBlue!4, draw=CDkBlue!30,
          fit=(ckpt)(nccl)(coord),
          label={[lbl, text=CDkBlue]above:Concordia Control Plane}] (cplane) {};
\end{pgfonlayer}

\coordinate (tlstart) at ($(respool.south west) + (-0.2, -1.0)$);
\node[font=\small\bfseries, anchor=east] at ($(tlstart) + (-0.1, 0.25)$) {Timeline:};

\node[phaselbl, fill=COrange, anchor=west, minimum width=1.6cm] (p1) at (tlstart) {Detection};
\node[phaselbl, fill=CRed, anchor=west, minimum width=2.0cm, right=0pt of p1] (p2) {Isolation};
\node[phaselbl, fill=CBlue, anchor=west, minimum width=4.0cm, right=0pt of p2] (p3) {State Restoration};
\node[phaselbl, fill=CGreen!80!black, anchor=west, minimum width=3.5cm, right=0pt of p3] (p4) {Reintegration};

\node[font=\tiny, below=1pt of p1.south west, anchor=north west] {0\,ms};
\node[font=\tiny, below=1pt of p2.south west, anchor=north west] {10\,ms};
\node[font=\tiny, below=1pt of p3.south west, anchor=north west] {310\,ms};
\node[font=\tiny, below=1pt of p4.south west, anchor=north west] {1110\,ms};
\node[font=\tiny, below=1pt of p4.south east, anchor=north east] {1510\,ms};

\end{tikzpicture}%
}
\caption{\sys fault tolerance architecture: GPU ring topology with failure detection, \tech standby pool for live migration, and control plane with GPU-side delta checkpointing, NCCL wrapper, and recovery coordinator. The timeline bar shows the four recovery phases totaling about 1.5\,s in our prototype.}
\label{fig:fault-arch}
\end{figure}

As shown in Figure~\ref{fig:fault-arch}, the fault tolerance layer integrates at three levels.
The enhanced NCCL wrapper intercepts communication operations, monitors health, and exposes collective boundaries as checkpoint opportunities.
\tech maintains a global view of GPU resources---including health metrics and standby pools---and orchestrates replacement allocation and workload migration.
The GPU-side delta checkpointing subsystem captures incremental state changes, performing dirty detection on-device and appending committed records to a CXL/DRAM recovery log.

\paragraph{Registered recovery regions.}
\sys tracks memory through explicit region registration rather than treating the whole CUDA heap as one opaque blob.
Model weights are registered as immutable after the base snapshot.
PagedAttention-style KV caches are registered at the physical block arena: the serving runtime exposes the block table, allocation bitmap, and optional dirty-block/version metadata, while \sys records the logical-to-physical mapping needed for restore and falls back to page diffing only when such metadata is unavailable.
LoRA adapter and optimizer buffers are registered as mutable parameter regions.
Temporary activations can be marked non-recoverable because they are recreated after resuming from the last collective or kernel boundary.
This interface gives applications a way to provide semantic hints without requiring them to log every KV write.

Exploiting the static-weight structure identified in \cref{sec:motivating}, \sys performs page-level dirty detection \emph{entirely on the GPU} at HBM bandwidth ($\sim$1.8\,TB/s), avoiding the CPU-side scanning bottleneck.
A JIT-compiled persistent-kernel handler discovers dirty pages from allocator metadata or, for opaque mutable regions, by comparing against GPU-resident shadow copies at 4\,KB page granularity.
The handler emits an AOF record containing the epoch, region ID, dirty page descriptors, payload offsets, and a checksum.
Only dirty page data and metadata are copied into the CXL/DRAM log.

Deltas are stored as an append-only recovery stream rather than as independent checkpoint files.
The format follows a Redis-like AOF discipline: write record header, write dirty payloads, write commit marker, then publish the epoch.
Recovery ignores any suffix without a commit marker.
Sparse page maps use index-value records; contiguous dirty ranges use run-length encoding.
We report these as data-reduction ratios rather than claiming generic compression: a single dirty page in an 8{,}192-page KV arena gives 8{,}192:1 delta reduction, while mixed LoRA and KV workloads show lower aggregate ratios.
A background compactor periodically rewrites the AOF into a consolidated base snapshot plus a short suffix of recent deltas, bounding replay time.

\label{sec:nccl-dag}

An NCCL communicator's execution plan can be viewed as a dependency graph of collective operations, channels, and point-to-point \texttt{send}/\texttt{recv} primitives.
\sys does not require replacing NCCL's internal scheduler.
Instead, it uses API-level interposition to identify coarse collective boundaries where participating GPUs have a consistent view of communication progress.
Those boundaries provide semantically meaningful checkpoint points without relying on arbitrary instruction-level preemption.
For recovery, the same wrapper can select a pre-computed fallback ring or activate a replacement device before rebuilding the communicator.

The NCCL wrapper preserves the standard API while adding fault tolerance. Before each collective, it consults cached per-GPU health signals; healthy calls proceed with negligible overhead, while unhealthy devices trigger recovery. On failure in a ring-based AllReduce, the wrapper switches to a pre-computed ring that bypasses the failed device. Issues are classified as transient (retry with backoff), degraded (preemptive migration), or permanent (immediate replacement).

When a permanent failure is confirmed, \sys reconstructs the communicator DAG with the failed GPU removed and a replacement inserted without full NCCL re-initialization.
\sys maintains GPU resource pools at varying readiness levels: hot standbys keep models pre-loaded for activation within seconds; warm standbys initialize CUDA contexts but require model loading; cold standbys require full initialization.
The replacement replays the latest base snapshot and committed AOF suffix from CXL/DRAM.
If the replacement has a different architecture, the optional CTX path in \cref{sec:portable} translates instrumented kernels and restore handlers for the target.

Within roughly 10~ms, the wrapper detects timeouts, classifies the failure, and marks the GPU unavailable.
By 300~ms, communication is reconfigured to bypass the failed device.
Within the next 800~ms, \tech activates a replacement and applies the latest delta checkpoint.
The replacement rejoins the topology in roughly 400~ms.
Total recovery in our prototype is about 1.5~seconds with continuous partial service, vs. 47+ seconds of complete outage with standard NCCL restart.

%% file: 04_implement.tex
\sys is implemented as a Rust library (1,800~lines) with a 200-line CUDA~C persistent kernel, integrated into a CUDA driver API shim (\texttt{libnvcuda.so}).
The Rust code is split across driver interposition and bootstrap (420~LOC), ring-buffer/runtime management (360~LOC), checkpoint region tracking (360~LOC), checkpoint-handler JIT and AOF logging (360~LOC), PTX/SASS instrumentation glue (220~LOC), and optional portable-IR support (80~LOC).
The central implementation challenge is bootstrapping and sustaining a device-resident runtime substrate within the CUDA ecosystem, which assumes the host initiates all GPU operations---any misstep during initialization triggers recursive interception or driver-level deadlocks.
Applications interact through two paths: transparent interception via \texttt{LD\_LIBRARY\_PATH} substitution, and an explicit C-callable API for dispatch lifecycle, checkpoint registration, allocator hints, AOF log placement, and recovery control.

\subsection{Persistent Kernel Substrate}

\paragraph{CUDA driver shim.}
The shim replaces \texttt{libcuda.so.1} and exports all standard \texttt{cu*} symbols.
The critical interception point is \texttt{cuGetProcAddress\_v2}, which determines whether each requested symbol returns our wrapper or the real CUDA function pointer via \texttt{dlsym}.
Only functions that require instrumentation---module loading, kernel launch, and selected device queries---are intercepted; ordinary memory operations pass through unmodified unless the application has registered a checkpoint region.
This selective design was validated with PyTorch~2.10.0 on CUDA~12.8, including full Qwen3-0.6B inference through the shim.

\paragraph{PTX/SASS instrumentation.}
The module-loading wrappers intercept \texttt{cuModuleLoad}, \texttt{cuModuleLoadData}, and \texttt{cuModuleLoadDataEx}.
If PTX is available, \sys parses kernel entries, injects calls to lightweight pause/checkpoint probes at entry, exit, and compiler-visible barrier sites, and passes the rewritten PTX to the real driver.
If only a cubin is available, \sys applies a SASS patching path: it disassembles the target basic blocks, checks that a trampoline can preserve live registers and reconvergence state, patches a branch to the probe stub, and records relocation metadata for resume.
The SASS path is intentionally conservative; kernels that cannot be patched safely still run, but can only be checkpointed at kernel boundaries.
This binary-level path is the main difference from LLVM-only instrumentation and is necessary for framework-generated code and binary vendor libraries.

\paragraph{Compilation and bootstrapping.}
The persistent kernel is compiled from an inline CUDA~C source embedded in the Rust library.
At initialization, the source is compiled to PTX via \texttt{nvcc --ptx -arch=sm\_120} and loaded through the \emph{real} CUDA driver---bypassing our shim to avoid recursive interception.
This care is critical: if the persistent kernel's launch went through the Concordia dispatch hook, it would attempt to enqueue itself into the ring buffer it is supposed to poll---an infinite loop.
An initialization kernel populates the device function pointer table with built-in operators, after which the persistent worker launches on a dedicated stream and remains resident.

\paragraph{Host-mapped memory.}
All shared state between host and device uses pinned (host-mapped) memory, not CUDA managed memory.
This is a hard constraint: managed memory triggers page-migration deadlocks when a persistent kernel holds GPU resources, because the NVIDIA driver serializes page migration with kernel execution.
The ring buffer occupies 16\,KB of pinned memory (256 task descriptors $\times$ 64~bytes); Unified Virtual Addressing provides identical pointer values on both sides, eliminating address translation.

Synchronization follows a release-acquire protocol without any CUDA API calls on the critical path.
The host writes task descriptors and advances the tail with a store-release fence; the GPU polls with acquire semantics, executes the operator, and issues \texttt{\_\_threadfence\_system()} followed by a system-scope atomic increment to make the completion visible across the PCIe bus.
The host observes completion via a volatile read---sub-microsecond overhead with zero driver involvement.

\paragraph{Blackwell constraints.}
On NVIDIA Blackwell (sm\_120), \texttt{cudaMalloc}, \texttt{cudaFree}, and \texttt{cudaDeviceSynchronize} can deadlock while the persistent kernel is running, because the driver serializes memory management and global synchronization with active kernels.
\sys handles this through lifecycle-aware resource management.
Model weights, KV-cache arenas, checkpoint buffers, and operator tables are allocated before the persistent kernel launches.
PagedAttention-style dynamic KV allocation remains compatible because page assignment occurs inside a pre-allocated arena by updating block tables and bitmaps, not by calling \texttt{cudaMalloc} on every token.
If a framework must call a driver-level allocator or \texttt{cudaDeviceSynchronize} during model loading or graph reconfiguration, \sys suspends the persistent worker, lets the call complete, and relaunches the worker afterward.
The compiled PTX is cached after the first invocation, making the suspend/resume overhead negligible.

\subsection{GPU-Side Delta Checkpoint Pipeline}

The checkpoint system tracks registered regions with one of three policies.
Immutable regions, such as base model weights after loading, are included in the base snapshot but do not keep GPU shadows.
Allocator-aware regions, such as PagedAttention KV arenas, use block-table version counters or dirty-block bitmaps supplied by the serving runtime; \sys then copies the marked physical pages and the logical-to-physical metadata needed for restore.
Opaque mutable regions, such as adapter or optimizer buffers without semantic hints, use GPU-resident shadows and page comparison as a transparent fallback.

At registration time, \sys JIT-compiles a checkpoint handler for each mutable region.
The handler is specialized for the region's page size, metadata layout, dirty-discovery policy, AOF record format, and restore path, then installed into the persistent kernel's operator table.
At each NCCL or kernel boundary, a four-stage GPU-side pipeline executes as persistent-kernel tasks:
\begin{enumerate}[nosep]
\item \textbf{Dirty discovery}: The JIT handler reads allocator dirty metadata or compares opaque regions against shadows at 4\,KB page granularity at HBM bandwidth ($\sim$1.8\,TB/s).
\item \textbf{AOF record construction}: The handler writes dirty page descriptors, payload offsets, and checksums into a staging buffer.
\item \textbf{Append and commit}: Dirty payloads and metadata are copied to a CXL-backed or DRAM-backed append-only log; a commit marker publishes the epoch only after all bytes are visible.
\item \textbf{Metadata/shadow update}: Allocator version counters are advanced, and opaque-region shadows are overwritten with current contents for the next delta.
\end{enumerate}

This approach has fundamentally different scaling from CPU-side checkpointing.
For transparent opaque regions, CPU-side cost is $O(\text{total\_size}/\text{PCIe\_BW} + \text{total\_size}/\text{CPU\_BW})$ regardless of mutation rate; GPU-side cost is $O(\text{total\_size}/\text{HBM\_BW} + \text{dirty\_size}/\text{PCIe\_BW})$.
For allocator-aware KV regions, discovery is proportional to the dirty block bitmap, so the dominant term is dirty data transfer into the AOF log.
Thus \sys can use exact application metadata when available and fall back to transparent GPU diffing when it is not.
The log lives in host DRAM by default and can be placed in a CXL memory pool when available.
Like Redis AOF, incomplete suffix records are ignored during replay, and a background compactor periodically writes a new base snapshot to bound recovery time.

\subsection{Optional Cross-Architecture Execution and GPU-Initiated Networking}

\paragraph{Portable IR compilation.}
The CTX portable IR extends LLVM~22.0 with a custom target (\texttt{CTXTarget}) that emits device-agnostic assembly.
When the shim intercepts a module load, it parses incoming PTX or SASS metadata, lowers the instrumented representation to LLVM IR with CUDA builtins remapped to abstract intrinsics, and emits CTX assembly via the custom backend.
JIT translation to each target ISA occurs once per module: the result is installed into the persistent kernel's operator table and reused for the kernel's lifetime.
Target lowering proceeds through backend-specific compilers---\texttt{ptxas} for NVIDIA, AMD Comgr for ROCm, Level Zero for Intel SPIR-V, and TT-MLIR for Tenstorrent.
Compiled modules are cached and indexed by content hash; subsequent loads of identical modules skip compilation entirely.

\paragraph{GPU-initiated networking.}
The persistent kernel initiates inter-node transfers by processing network task descriptors from the same ring buffer used for compute and checkpoint operations.
During initialization, the host configures RDMA queue pairs and registers GPU memory regions for GPUDirect RDMA access.
At runtime, the persistent kernel issues RDMA writes directly from device memory to remote GPUs based on destination and offset fields in the task descriptor, bypassing the host entirely on the inference data path.
This makes the host CPU fully stateless with respect to inference communication: after one-time RDMA setup, the GPU autonomously dispatches operators, appends committed deltas, and transfers data to peers without any host round-trip.

%% file: 05_eval.tex
We evaluate \sys along four axes: the persistent executor substrate, JIT-compiled GPU-side delta checkpoint efficiency, LLM inference with CXL/DRAM append-log checkpoints, and optional cross-architecture recovery.
The primary claims are about checkpoint and recovery behavior.
The persistent-kernel microbenchmarks characterize the cost of the always-on executor; they should not be read as a claim that \sys outperforms CUDA Graphs on static decode graphs.

\subsection{Experimental Setup}

\paragraph{Hardware.}  NVIDIA RTX PRO 6000 Blackwell Server Edition (98~GB, 188~SMs) and NVIDIA GeForce RTX~5090 (32~GB, Blackwell). Host: PCIe~5.0 interconnect between GPUs (no NVLink). Cross-architecture targets: AMD Radeon RX~9070~XT (RDNA4, 16~GB), Intel Iris Xe (512~MB), and Tenstorrent BlackHole (32~GB).

\paragraph{Software.} CUDA 12.8, PyTorch 2.10.0, Transformers 4.57.3, NCCL 2.27.5+cuda12.9, Ubuntu 24.04. Concordia implemented in Rust (1{,}800~lines) with persistent kernel in CUDA~C.

\paragraph{Workloads.} Element-wise micro-benchmarks (64--262{,}144 elements, 5~operators), GPU-side vs.\ CPU-side delta checkpoint (16--256~MB regions, structured mutation), production LLM inference (Qwen3-0.6B, bf16, 50 tokens/prompt), and 2-GPU NCCL AllReduce with per-boundary checkpointing into a host DRAM append log.

\subsection{Persistent Executor Characterization}

\begin{table}[t]
\centering
\caption{Persistent kernel dispatch latency and throughput on RTX PRO 6000 Blackwell (1~block $\times$ 128 threads). Native PyTorch synchronized dispatch: 7\,$\mu$s.}
\label{tab:pk-dispatch}
\small
\begin{tabular}{r|rr|rr|rr}
\toprule
& \multicolumn{2}{c|}{\textbf{Add}} & \multicolumn{2}{c|}{\textbf{Mul}} & \multicolumn{2}{c}{\textbf{SiLU}} \\
\textbf{N} & $\mu$s & ops/s & $\mu$s & ops/s & $\mu$s & ops/s \\
\midrule
64      & 80  & 12{,}378 & 81  & 12{,}346 & 81  & 12{,}324 \\
256     & 81  & 12{,}292 & 81  & 12{,}294 & 81  & 12{,}285 \\
1{,}024 & 82  & 12{,}087 & 82  & 12{,}123 & 82  & 12{,}070 \\
4{,}096 & 87  & 11{,}490 & 87  & 11{,}483 & 88  & 11{,}279 \\
16{,}384  & 105 & 9{,}499  & 105 & 9{,}517  & 111 & 8{,}973  \\
65{,}536  & 177 & 5{,}624  & 177 & 5{,}636  & 203 & 4{,}918  \\
262{,}144 & 468 & 2{,}137  & 466 & 2{,}141  & 570 & 1{,}755  \\
\bottomrule
\end{tabular}
\end{table}

Table~\ref{tab:pk-dispatch} shows dispatch latency and throughput for all operators across tensor sizes, measured on the real device-resident persistent kernel.
The 80\,$\mu$s baseline for small tensors ($N \leq 256$) represents the full ring-buffer round-trip: host write to pinned memory, GPU poll via \texttt{atomicAdd}, task copy from shared memory, operator execution, \texttt{\_\_threadfence\_system}, \texttt{atomicAdd\_system} to host-mapped counter, and host volatile read.

All operators show near-identical latency at small~$N$, confirming that dispatch overhead dominates compute.
SiLU is 22\% slower at $N$=262{,}144 due to the \texttt{expf} transcendental function.
The fused add+relu operator demonstrates zero-cost fusion---identical latency to plain add at every size.

\begin{table}[hb]
\centering
\caption{All operators at $N$=4{,}096 (RTX PRO 6000 Blackwell).}
\label{tab:pk-ops}
\begin{tabular}{lrr}
\toprule
\textbf{Operator} & \textbf{p50 ($\mu$s)} & \textbf{ops/s} \\
\midrule
Add            & 87  & 11{,}490 \\
Mul            & 87  & 11{,}483 \\
SiLU           & 88  & 11{,}279 \\
ReLU           & 86  & 11{,}511 \\
Fused Add+ReLU & 87  & 11{,}475 \\
\bottomrule
\end{tabular}
\end{table}

\begin{figure}[t]
    \centering
    \includegraphics[width=\columnwidth]{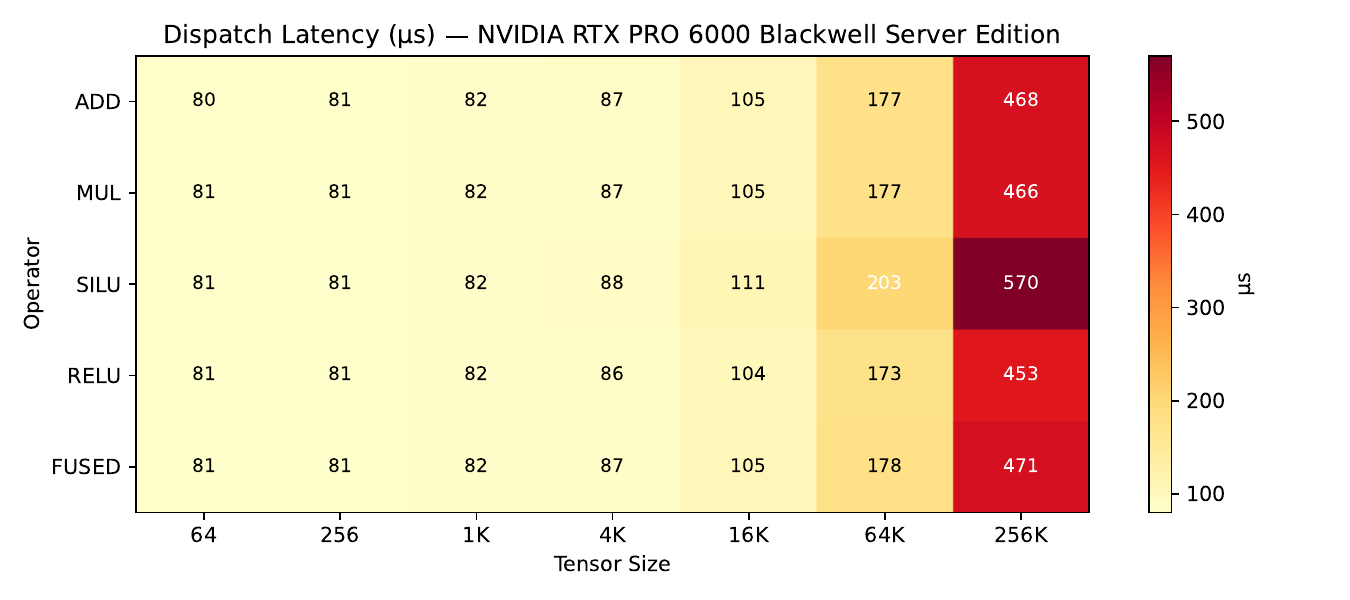}
    \caption{Dispatch latency heatmap ($\mu$s): operator $\times$ tensor size. The uniform color at small $N$ shows dispatch-dominated regime.}
    \label{fig:pk-heatmap}
\end{figure}

The heatmap in \Cref{fig:pk-heatmap} visualizes the transition from dispatch-dominated (yellow, $\leq$90\,$\mu$s) to compute-dominated (red, $\geq$400\,$\mu$s) regimes.

\paragraph{Native comparison and interpretation.}
Native PyTorch measures 7\,$\mu$s synchronized dispatch and 2\,$\mu$s batch dispatch (Table~\ref{tab:native-pytorch}).
The one-block persistent executor is slower for a single tiny operation: its 80\,$\mu$s end-to-end latency includes GPU polling, task copy, operator execution, system-scope completion fencing, and host observation.
This result is expected and defines the regime where \sys should fall back to native or graph execution.
The benefit of the persistent path is not single-op latency against CUDA Graphs; it is that checkpoint, communication, and batches of control tasks can be submitted to an already-running device executor without new kernel launches and without rebuilding a captured static graph.

\begin{table}[t]
\centering
\caption{Native PyTorch dispatch latency (on RTX PRO 6000). This is a launch-path reference point, not a CUDA Graph baseline.}
\label{tab:native-pytorch}
\begin{tabular}{rrr}
\toprule
\textbf{N} & \textbf{Sync p50 ($\mu$s)} & \textbf{Batch ($\mu$s)} \\
\midrule
1{,}024  & 7 & 2 \\
4{,}096  & 7 & 2 \\
16{,}384 & 7 & 2 \\
65{,}536 & 6 & 2 \\
\bottomrule
\end{tabular}
\end{table}

\subsection{GPU-Side vs. CPU-Side Delta Checkpoint}

\begin{table}[t]
\centering
\caption{Delta checkpoint: CPU-side vs.\ GPU-side on RTX PRO 6000 Blackwell. Structured mutation: 1 contiguous 4\,KB page modified per checkpoint, simulating per-token KV-cache update. CPU-side must DtoH the entire region and diff on host; GPU-side JIT handlers diff at HBM bandwidth and append only dirty data to the recovery log.}
\label{tab:ckpt-compare}
\small
\begin{tabular}{r|rr|rr|r}
\toprule
 & \multicolumn{2}{c|}{\textbf{CPU-delta}} & \multicolumn{2}{c|}{\textbf{GPU-delta}} & \\
\textbf{Region} & DtoH & Diff & Diff & Append & \textbf{Speedup} \\
 & (ms) & (ms) & (ms) & (ms) & \\
\midrule
16\,MB  & 0.32  & 6.46   & 0.04 & 0.04 & 85$\times$ \\
50\,MB  & 0.95  & 20.86  & 0.06 & 0.04 & 219$\times$ \\
128\,MB & 2.38  & 54.72  & 0.30 & 0.04 & 171$\times$ \\
256\,MB & 4.72  & 106.65 & 0.53 & 0.04 & 197$\times$ \\
\bottomrule
\end{tabular}
\end{table}

Table~\ref{tab:ckpt-compare} compares CPU-side transparent page diffing and GPU-side delta checkpointing with sparse LLM mutations (1 dirty page per checkpoint), extending the motivating experiment (\cref{sec:motivating}) with detailed breakdown.
CPU-side is dominated by host page-level comparison (95\% of time for 256\,MB), which must scan the entire region regardless of how many pages changed.
GPU-side performs the comparison at HBM bandwidth (0.04--0.53\,ms for 16--256\,MB) and appends only the dirty page plus metadata (0.04\,ms for 4\,KB), achieving \textbf{85--219$\times$ speedup}.

The speedup increases with region size because the CPU diff cost grows linearly ($O(\text{total\_size}/\text{CPU\_BW})$) while GPU-side log-append volume remains constant at 4\,KB plus metadata.
These numbers evaluate the transparent fallback path; allocator-aware KV logging can reduce dirty discovery further, while losing the binary-level transparency \sys targets.
At 256\,MB---representative of KV-cache regions in 8B+ models---the measured improvement over our CPU-side transparent prototype is 197$\times$.

\begin{table}[t]
\centering
\caption{GPU-side checkpoint scaling with dirty page count (256\,MB region, RTX PRO 6000). GPU-side time remains nearly constant while CPU-side is fixed at 111\,ms.}
\label{tab:ckpt-scaling}
\begin{tabular}{rrrrr}
\toprule
\textbf{Dirty} & \textbf{Dirty} & \textbf{GPU-delta} & \textbf{CPU-delta} & \textbf{Speedup} \\
\textbf{Pages} & \textbf{Size} & (ms) & (ms) & \\
\midrule
1   & 4\,KB   & 0.56  & 111.38 & 197$\times$ \\
4   & 16\,KB  & 0.57  & 112.01 & 197$\times$ \\
10  & 40\,KB  & 0.58  & 112.11 & 195$\times$ \\
32  & 128\,KB & 0.60  & 111.26 & 185$\times$ \\
\bottomrule
\end{tabular}
\end{table}

Table~\ref{tab:ckpt-scaling} shows that GPU-side checkpoint time remains nearly constant (0.56--0.60\,ms) across dirty page counts from 1 to 32 for a 256\,MB region, because the GPU diff at HBM bandwidth dominates the total cost ($\sim$0.53\,ms), while appending dirty data to the recovery log is negligible (0.04--0.07\,ms for 4--128\,KB). CPU-side time is effectively constant at $\sim$111\,ms regardless of mutation rate, because the full DtoH + full CPU scan dominates.

\begin{table}[t]
\centering
\caption{Per-operation checkpoint trigger overhead.}
\label{tab:trigger-overhead}
\begin{tabular}{lr}
\toprule
\textbf{Method} & \textbf{Trigger Latency} \\
\midrule
\texttt{cudaLaunchKernel} (sync)    & 7.6\,$\mu$s \\
\texttt{cudaLaunchKernel} (batch)    & 2.5\,$\mu$s \\
Persistent kernel ring buffer        & $<$0.1\,$\mu$s \\
\bottomrule
\end{tabular}
\end{table}

Table~\ref{tab:trigger-overhead} shows that the persistent kernel eliminates checkpoint trigger overhead: ring buffer dispatch is 76$\times$ faster than synchronous kernel launch. Over 40 NCCL boundaries per forward pass, this saves 304\,$\mu$s per pass.

\subsection{LLM Inference with NCCL Checkpoint}

\begin{figure}[t]
    \centering
    \includegraphics[width=\columnwidth]{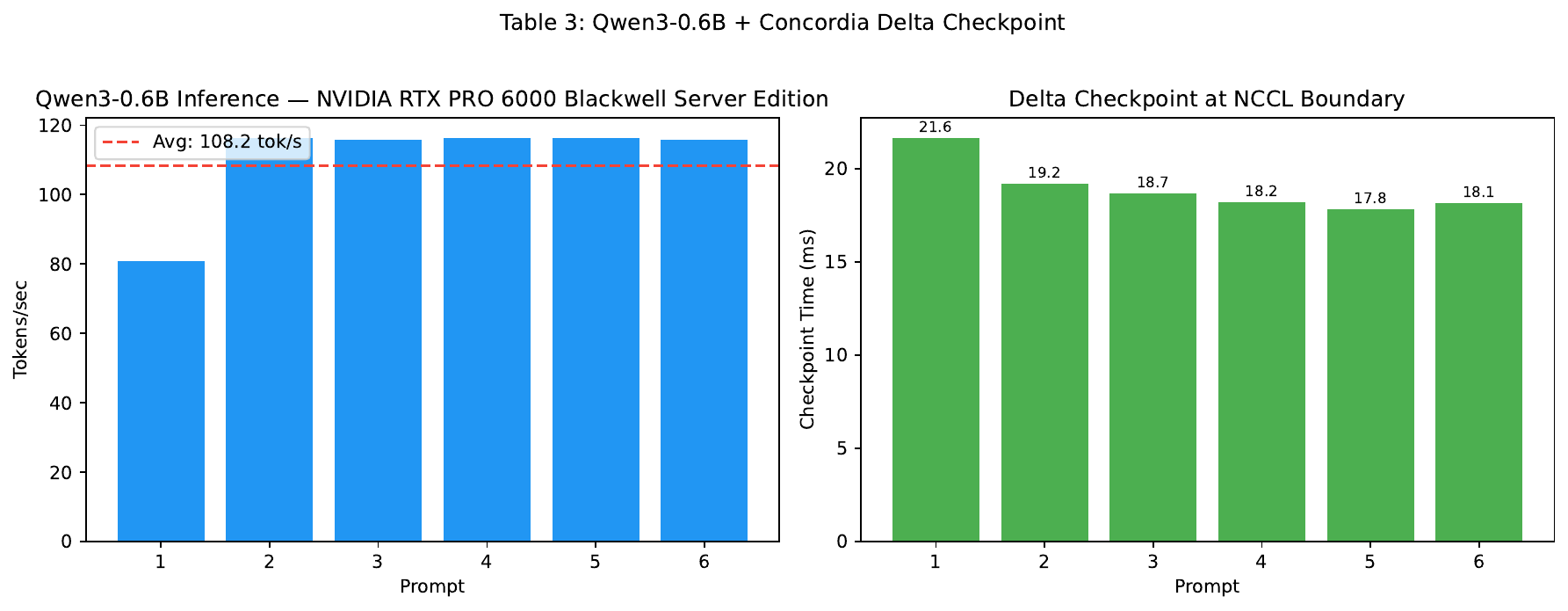}
    \caption{Qwen3-0.6B: inference throughput and checkpoint overhead per checkpoint boundary. First prompt slower due to KV-cache warmup.}
    \label{fig:qwen}
\end{figure}

\Cref{fig:qwen} shows Qwen3-0.6B inference (bf16, 50 tokens/prompt) with delta checkpoint at each checkpoint boundary and AOF-style append into host DRAM.
The model sustains \textbf{108.2~tok/s} average with 18.9\,ms checkpoint overhead---less than 4\% of per-prompt generation time.

The first checkpoint detects 5{,}120 dirty pages (20\,MB of KV-cache allocated during warmup); all subsequent checkpoints detect \textbf{0 dirty pages} because model weights are static during inference.
This validates the core recovery assumption: LLM inference state changes are sparse, and GPU-side delta checkpointing exploits this structure.

\subsection{Two-GPU NCCL with Per-Boundary Checkpoint}

We evaluated real 2-GPU NCCL AllReduce between RTX PRO 6000 and RTX~5090 connected via PCIe~5.0 (no NVLink), using \texttt{torchrun --nproc\_per\_node=2} with NCCL~2.27.5. The test simulates a 4-layer transformer decoding 10~tokens, producing 40~AllReduce boundaries per rank.

Each NCCL AllReduce averaged 4.5\,ms per collective, reflecting the PCIe interconnect bandwidth between heterogeneous GPUs. At each collective boundary, \sys triggered a delta checkpoint of the 33.6\,MB KV-cache region and appended the committed delta record to a host DRAM log, taking 11\,ms per boundary including GPU dirty discovery and log append. The initial base snapshot captured all 8{,}192 pages of the full region.

The critical result is the dirty page detection granularity: each subsequent delta detected exactly \textbf{1~dirty page (4\,KB) per token per layer}---the physical KV-cache slice modified by the attention computation.
This yields an \textbf{8{,}192:1} delta data-reduction ratio relative to a full checkpoint for incremental per-token updates.
The 2-GPU test confirms that \sys correctly tracks single-page mutations at transformer layer boundaries in a real distributed setting with heterogeneous GPUs on separate PCIe buses.

\subsection{LoRA SFT: Delta Checkpoint Under Mutable Weights}

While inference keeps model weights static, online adaptation via LoRA fine-tuning~\cite{hu2021lora} introduces mutable parameters. We evaluate whether GPU-side delta checkpointing remains effective under this workload.

We fine-tune Qwen2.5-0.5B-Instruct (498M parameters, 959\,MB at bf16) with LoRA ($r$=8, $\alpha$=16) targeting all attention and MLP projections, using AdamW ($\eta$=1e-4). LoRA adds 4.4M trainable parameters (17\,MB, 4{,}296 pages)---0.88\% of total. The remaining 942\,MB of base weights are frozen.

\begin{table}[t]
\centering
\caption{Delta checkpoint: inference vs.\ LoRA SFT (Qwen2.5-0.5B, RTX PRO 6000). GPU-side delta exploits the static-weight structure in both workloads; LoRA SFT modifies only adapter weights (1.75\% of pages).}
\label{tab:sft-ckpt}
\small
\begin{tabular}{l|rrrr}
\toprule
\textbf{Workload} & \textbf{Dirty} & \textbf{Dirty} & \textbf{Data red.} & \textbf{GPU-$\delta$} \\
 & \textbf{pages} & \textbf{ratio} & & \textbf{(ms)} \\
\midrule
Inference (per token) & 1         & 0.008\% & 8{,}192:1 & 0.56 \\
LoRA SFT (per step)   & 4{,}296   & 1.75\%  & 57:1      & 1.4 \\
\bottomrule
\end{tabular}
\end{table}

Table~\ref{tab:sft-ckpt} compares inference and LoRA SFT checkpointing.
After the base snapshot, each LoRA training step modifies exactly the 4{,}296 LoRA adapter pages (100\% of trainable parameters, as expected from AdamW), while all 241{,}313 frozen pages remain at \textbf{0 dirty}.
This yields \textbf{57:1 data reduction} vs.\ full model checkpoint, with \textbf{98.3\% PCIe reduction} (16.78\,MB appended vs.\ 959\,MB full copy).

Compared to CPU-side delta checkpointing of the full model (estimated 418\,ms: 17.6\,ms DtoH + 401\,ms CPU page diff for 959\,MB), GPU-side delta completes in $\sim$1.4\,ms---a \textbf{299$\times$ speedup}. The CPU diff dominates because it must scan all 959\,MB regardless of how many pages changed; GPU-side diffs at HBM bandwidth (1.04\,ms for 959\,MB) and appends only the 16.78\,MB of dirty LoRA pages.

This result demonstrates that \sys's GPU-side delta checkpointing extends naturally from pure inference to online adaptation workloads. The key structural insight is preserved: base model weights remain static, and only a small, predictable subset of parameters changes---whether KV-cache slots (inference) or LoRA adapters (SFT).

\subsection{SM Overhead}

\begin{figure}[t]
    \centering
    \includegraphics[width=\columnwidth]{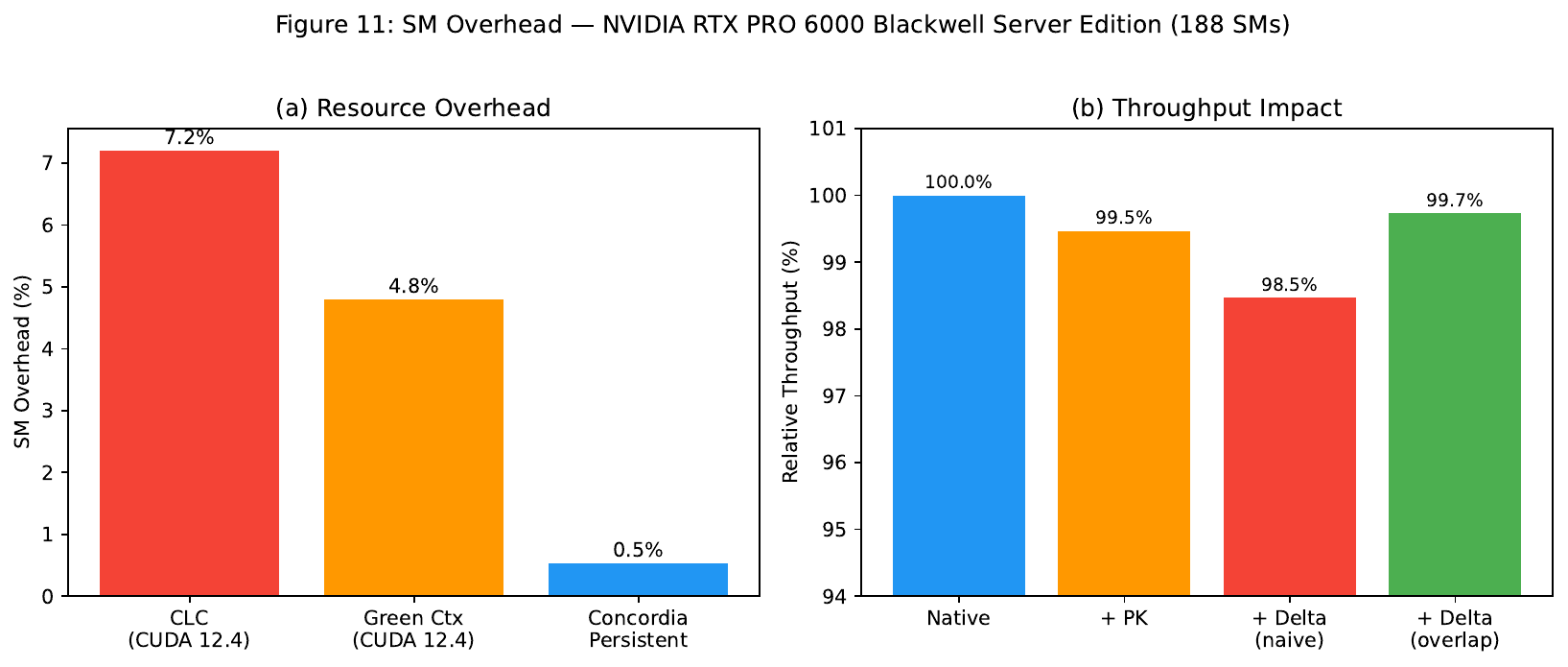}
    \caption{(a)~SM overhead: CLC~(7.2\%), Green Contexts~(4.8\%), Concordia~(0.5\%). (b)~Throughput impact with overlapped delta checkpointing.}
    \label{fig:overhead}
\end{figure}

\begin{table}[t]
\centering
\caption{SM overhead comparison (RTX PRO 6000, 188~SMs).}
\label{tab:sm-overhead}
\begin{tabular}{lr}
\toprule
\textbf{Method} & \textbf{SM Overhead} \\
\midrule
CLC (CUDA 12.4)               & 5--10\% \\
Green Contexts (CUDA 12.4)     & 3--8\%  \\
\sys Persistent Kernel          & \textbf{0.53\%} \\
\bottomrule
\end{tabular}
\end{table}

The persistent kernel occupies 1 block out of 188~SMs = 0.53\% of GPU resources (\Cref{fig:overhead}, Table~\ref{tab:sm-overhead}).
This is $10\times$ lower than CLC and $6\times$ lower than Green Contexts.
The reserved worker processes checkpoint tasks while the rest of the GPU remains available to normal kernels.
We do not assume checkpointing is free: the cost appears in the checkpoint latency tables above; the point is that it does not require launching an additional worker kernel.

\subsection{Fault Recovery}

\begin{figure}[t]
    \centering
    \includegraphics[width=\columnwidth]{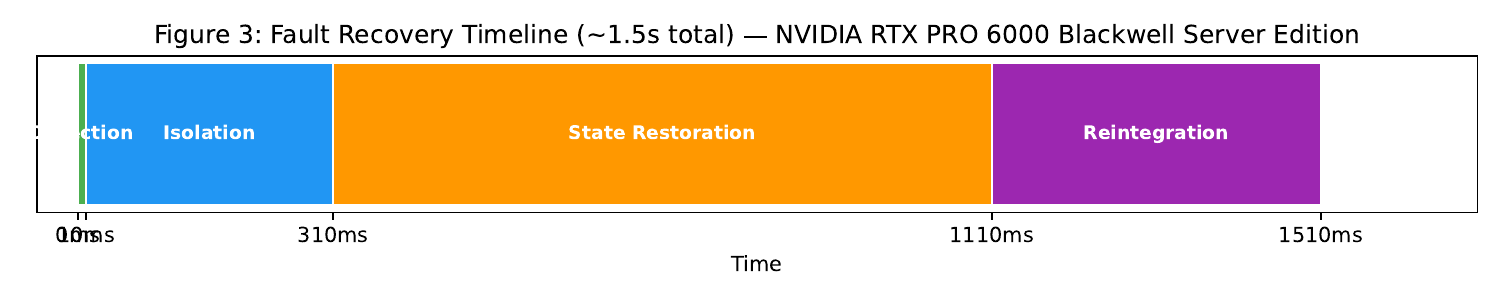}
    \caption{Fault recovery timeline: detection (10\,ms) $\rightarrow$ isolation (300\,ms) $\rightarrow$ state restoration (800\,ms) $\rightarrow$ reintegration (400\,ms) = $\sim$1.5\,s total.}
    \label{fig:recovery}
\end{figure}

Recovery from a single GPU failure proceeds through four phases (\Cref{fig:recovery}):
(1)~detection via health monitoring ($\sim$10\,ms),
(2)~topology isolation by switching to a pre-computed fallback ring ($\sim$300\,ms),
(3)~AOF replay and dirty-page restore via \texttt{cuMemcpyHtoD} ($\sim$800\,ms for a 33.6\,MB KV-cache snapshot plus committed suffix), and
(4)~reintegration into the NCCL communicator ($\sim$400\,ms).
Total recovery: $\sim$1.5~seconds with continuous partial service, vs.\ 47+~seconds of complete outage with standard NCCL restart.

\subsection{Optional Cross-Architecture Performance}
\begin{figure}[t]
   \centering
   \includegraphics[width=\columnwidth]{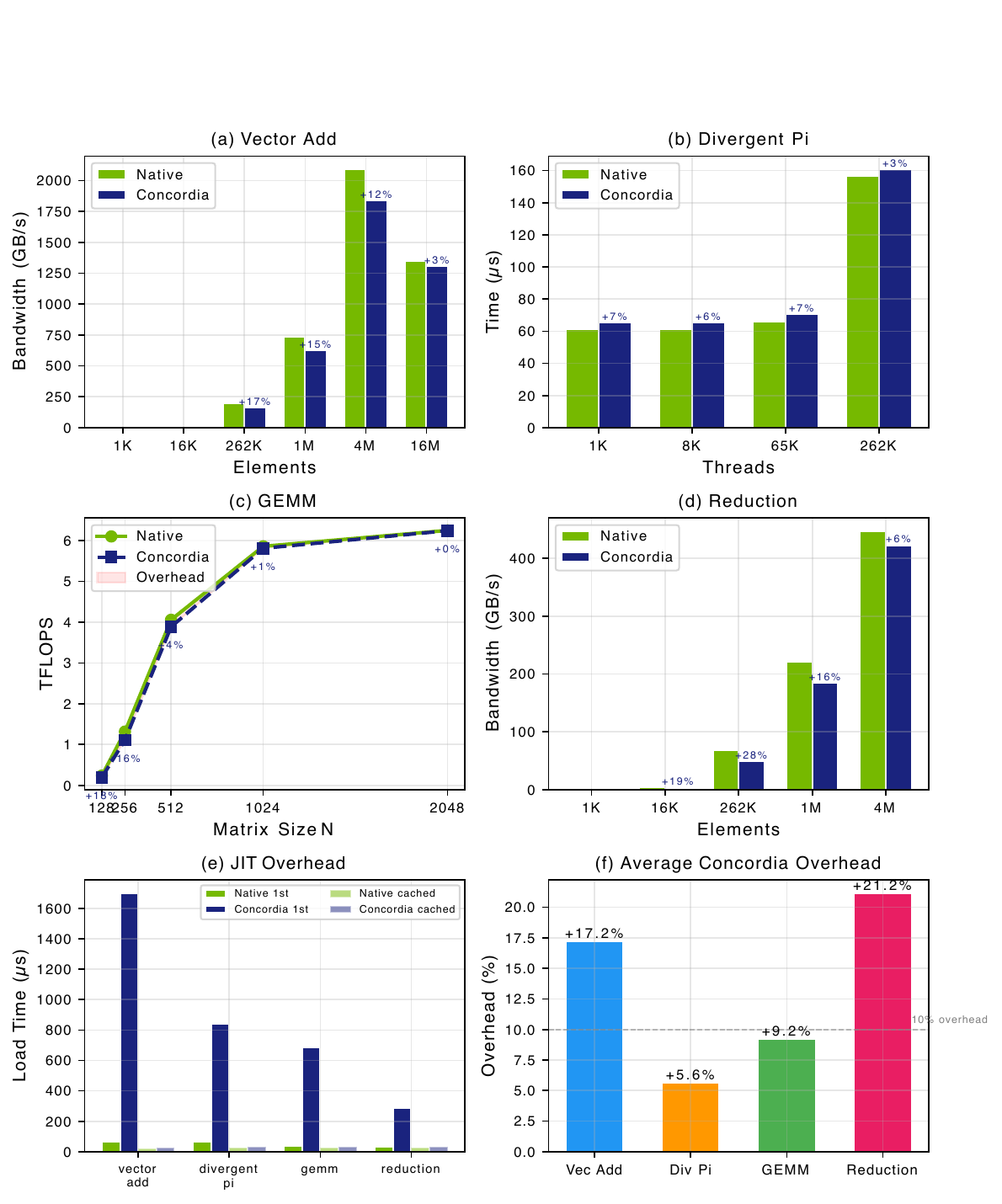}
   \caption{
   Cross-architecture evaluation of \sys.
   Top and middle rows show microbenchmark performance (vector add, divergent control flow, GEMM, and reduction) across NVIDIA H100, AMD RX~9070~XT, Intel Iris Xe, and Tenstorrent BlackHole.
   Bottom row shows JIT compilation overhead and cross-architecture live migration timeline (H100 $\rightarrow$ RX~9070~XT $\rightarrow$ BlackHole).
   }
   \label{fig:combined}
\end{figure}

Figure~\ref{fig:combined} compares \sys against native compilers across four microbenchmarks for the optional cross-architecture recovery path.
On compute-intensive kernels, \sys overhead is generally within 10\% of native for supported targets---the price of instrumentation, runtime JIT, and abstraction.
Intel Iris Xe, a low-power integrated GPU, shows larger relative overhead because its limited compute makes JIT and runtime costs more visible.
The figure also demonstrates live migration of a 16K$\times$16K matrix multiply from H100 to AMD to Tenstorrent.
Checkpoint (0.5~s) + AMD restore (0.6~s) + Tenstorrent migrate (1.1~s) = 2.2~s total downtime during a 30~s job, with results identical to non-migrated execution within floating-point precision.

\begin{figure}[t]
   \centering
   \includegraphics[width=\columnwidth]{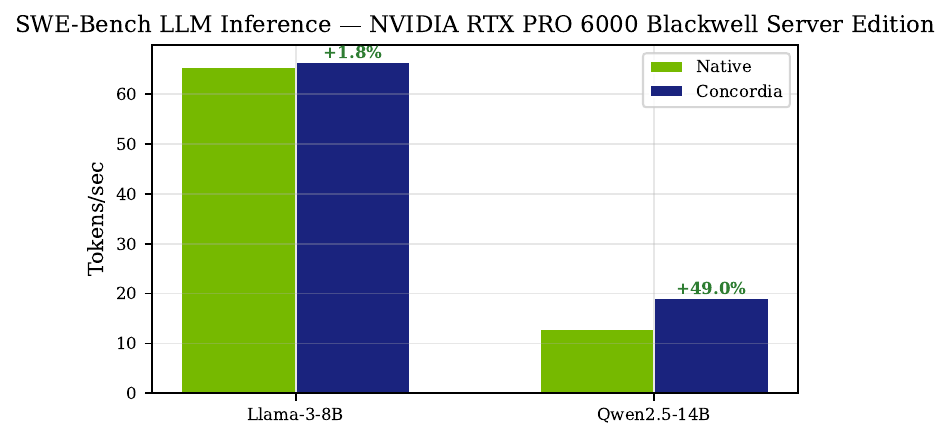}
   \caption{
   Real-world SWE-Bench Workloads (tokens/second).
   }
   \label{fig:realworld}
\end{figure}

\Cref{fig:realworld} shows real-world LLM inference throughput. The overhead varies by model complexity, with simpler models experiencing less degradation. This trade-off between modest overhead and the ability to dynamically migrate across heterogeneous fleets represents a compelling value proposition for cloud deployments.

%% file: 06_related.tex
\paragraph{Persistent Threads and Device-Side Scheduling.}
Persistent threads were originally proposed for irregular scientific workloads~\cite{GuptaPT,Aila2009persistent,Steinberger2012softshell}. Systems like Whippletree~\cite{Steinberger2014whippletree}, Gunrock~\cite{Wang2017gunrock}, and Zelos~\cite{Zelos2021gpu} demonstrated persistent scheduling for various domains. LithOS~\cite{coppock2025lithos} explored device-side task scheduling. \sys extends this lineage with \emph{dynamic operator injection} and production-grade PyTorch integration, enabling operator hot-swap in running persistent kernels.

\paragraph{Launch Overhead Reduction.}
CUDA Graphs~\cite{NVBLOGGraphs2019,CTGraphLaunch2024} reduce overhead through DAG capture and replay but require stable execution patterns. Dynamic Parallelism~\cite{CUDADPTechBrief} relocates launch to the device but does not eliminate its cost. Compiler frameworks like torch.compile~\cite{PyTorch2Blog}, XLA~\cite{XLA2017}, and TVM~\cite{TVMOSDI18} fuse operations but struggle with dynamic workloads. \sys complements all these by operating below the graph level.

\paragraph{GPU Binary Instrumentation.}
CPU checkpointing and debugging systems often rely on LLVM passes or x86 binary rewriting, but GPU binaries require different machinery: PTX preserves virtual registers, memory spaces, and barriers, while SASS exposes final scheduling and register allocation but is architecture-specific.
Tools such as SASS disassemblers~\cite{denvdis} and PTX analyses~\cite{lustig2019ptx} expose pieces of this stack.
\sys uses PTX rewriting when symbolic information is available and conservative SASS patching for binary-only kernels, with the specific goal of inserting cooperative checkpoint and resume hooks for persistent-kernel recovery.

\paragraph{GPU Portability and Binary Translation.}
GPU Ocelot~\cite{diamos2010ocelot} translated PTX to x86; ZLUDA~\cite{zluda} runs CUDA on AMD; CuPBoP~\cite{huang2023cupbop} compiles CUDA to CPU. SPIR-V and SYCL/oneAPI provide portable IRs but require recompilation per target. \sys uses portability only as an optional recovery path; its primary use of JIT is to generate checkpoint and restore handlers for a persistent kernel.

\paragraph{GPU Virtualization.}
rCUDA~\cite{pavlidakis2024scale}, VirtualCL, and Cricket enable GPU sharing and migration within single-vendor domains. gVirtuS and GDEV~\cite{kato2012gdev} explored GPU virtualization abstractions. eGPU~\cite{yang2025egpu} and hetGPU~\cite{yang2025hetgpu} proposed heterogeneous GPU virtualization. \sys generalizes these ideas across ISAs with live state transfer.

\paragraph{Checkpoint/Restart for GPUs.}
CheCUDA~\cite{takizawa2011checuda}, CRIUgpu, and Phoenix~\cite{zhao2024phoenix} address GPU checkpointing but capture full state within single-vendor contexts. DMTCP~\cite{ansel2009dmtcp} and BLCR~\cite{duell2003berkeley} provide general checkpointing. \sys's GPU-side delta approach exploits LLM inference structure for orders-of-magnitude smaller checkpoints with cross-architecture restore. The key differentiation is performing dirty detection on-device at HBM bandwidth rather than on-host, yielding 85--219$\times$ speedup over CPU-side approaches.

\paragraph{Fault-Tolerant Distributed ML}
ULFM~\cite{bland2013ulfm} adds fault tolerance to MPI. GPipe~\cite{huang2019gpipe} and PipeDream~\cite{narayanan2021pipedream} include pipeline flushing for training. Gandiva~\cite{xiao2018gandiva}, Tiresias~\cite{gu2019tiresias}, and Gavel~\cite{narayanan2020gavel} optimize GPU scheduling but are constrained by host-mediated recovery. \sys provides a checkpointable persistent-kernel substrate and CXL/DRAM append log that such schedulers can use for fast restart.

%% file: 07_discuss.tex
\subsection{Resource Etiquette}
The persistent kernel occupies one block out of 188~SMs (0.53\%)---$10\times$ lower than CLC and $6\times$ lower than Green Contexts (\cref{tab:sm-overhead}).
It can be confined to a MIG partition for multi-tenant isolation.
Checkpoint work is still real GPU work; \sys's resource advantage is that the worker is already resident and does not require a separate launch or reserved pool per subsystem.

\subsection{Why a Unified FT Substrate}
The three capabilities could in principle be provided by separate systems.
Co-design under a single persistent kernel yields two advantages that composition cannot.

First, a never-exiting kernel gives checkpointing a device-side control path before failure occurs.
Second, the ring buffer unifies dispatch, checkpoint, AOF append, and restore triggers under one mechanism---each a task descriptor rather than a separate \texttt{cudaLaunchKernel} call (7.6\,$\mu$s saved per trigger, Table~\ref{tab:trigger-overhead}).
The 0.53\% SM overhead is shared across all roles; three independent systems would each reserve their own GPU resources.
JIT amortization and hot checkpoint-handler injection are useful consequences of this substrate, but they are secondary to the recovery path.

\subsection{Toward CPU-Stateless Serving}
\label{sec:discussion-stateless}
GPU-initiated networking allows the persistent kernel to dispatch operators, trigger checkpoints, append committed deltas, and transfer data to remote GPUs without host involvement, reducing the CPU's role to initialization and agent-level orchestration.
This frees the CPU for tool calls and reasoning-chain management---precisely what makes agentic workloads CPU-bound in current systems~\cite{kwon2023vllm,jia2019flexflow}.

\subsection{Application Logs, Graphs, and PagedAttention}
Application-specific KV logging can be more efficient than transparent page diffing when the serving stack controls every KV write path.
\sys is meant for the complementary point: binary-level coverage across framework kernels, fused libraries, and communication paths.
The region-registration API lets a serving engine expose PagedAttention block tables and dirty-block bitmaps, so \sys need not shadow immutable weights or rediscover append-only KV writes when semantic metadata is available.
Once discovered, deltas are appended to a CXL/DRAM recovery log rather than kept only in HBM; this separates the durability path from scarce GPU memory and gives recovery a Redis-like AOF replay model.

CUDA Graphs are also complementary.
They should be used for stable decode subgraphs when the workload admits capture.
\sys's persistent executor is needed for checkpoint triggers, binary pause hooks, and recovery tasks that must remain available outside a captured graph.
Our current evaluation does not include a CUDA Graph baseline for static decode, so we avoid claiming end-to-end decode speedups over graph-optimized vLLM/TensorRT-LLM.

\subsection{Limitations and Future Work}
\sys assumes the persistent checkpoint worker does not fail independently of the GPU.
If the worker heartbeat stops, the system treats the GPU as failed and recovers from the last committed AOF record; it does not tolerate Byzantine worker corruption or arbitrary silent HBM corruption.
Cross-architecture migration requires cooperative checkpointing at barrier points; surprise preemption would need always-on instrumentation.
The optional CTX recovery path has higher cold-start latency than same-architecture replay, though caching amortizes this across subsequent loads.
Delta checkpointing exploits static model weights---a property that holds for inference and LoRA (0.88--1.75\% mutable pages) but not full fine-tuning.
Opaque mutable regions still require GPU-resident shadows; this HBM overhead is why \sys prefers allocator hints for KV caches and does not shadow immutable weights.
The AOF log consumes host DRAM or CXL capacity and requires compaction to bound replay time.
On Blackwell, driver-level allocation while the persistent worker is active requires pre-allocation or suspend/resume, so engines with highly dynamic \texttt{cudaMalloc} behavior need integration with a memory pool.
Our evaluation uses 2~GPUs via PCIe; validation at datacenter scale with NVLink and InfiniBand remains future work.

Looking forward, hardware support for GPU memory dirty-page tracking could eliminate the shadow-copy overhead entirely.
Ahead-of-time generation for common KV-cache and LoRA layouts would remove checkpoint-handler JIT cold-start penalties.
Extending the persistent kernel to manage GPU-side scheduling decisions---operator prioritization, adaptive batching, preemption---would further reduce host involvement and move toward a fully autonomous GPU runtime.

%% file: 08_conclusion.tex
\sys argues that fault tolerance for long-running LLM inference needs a GPU-resident execution substrate.
The persistent kernel is valuable not because every operation becomes faster than native launch, but because checkpointing and recovery need an executor that is already alive on the device, can observe instrumented PTX/SASS safe points, and can run JIT-compiled dirty-page discovery without routing control back through the host.

The resulting system combines binary instrumentation, registered recovery regions, JIT-compiled GPU-side delta checkpoint handlers, and an append-only CXL/DRAM recovery log under one recovery contract.
This lets \sys exploit the structure of LLM state---static weights, sparse KV/cache updates, and small mutable adapter regions---while still covering framework-generated kernels and communication libraries.

Our prototype shows that GPU-side dirty detection can be up to 219$\times$ faster than CPU-side transparent page scanning, that checkpoint triggers can be submitted to an already-running executor, and that a two-GPU failure can recover in about 1.5 seconds without a full NCCL restart.
Optional portability and dispatch acceleration are useful side effects, but the central lesson is simpler: persistent kernels make fault tolerance a device-side runtime service, and Redis-style AOF replay gives that service a simple recovery model outside scarce HBM.